\documentclass[acmsmall]{acmart}

\acmJournal{PACMPL}
\acmVolume{1}
\acmNumber{CONF} 
\acmArticle{1}
\acmYear{2018}
\acmMonth{1}
\acmDOI{} 
\startPage{1}

\setcopyright{none}

\newcommand{\eat}[1]{}
\usepackage{footnote}
\usepackage{longtable}
\usepackage{ltablex}
\usepackage{ltablex}
\makesavenoteenv{tabular}
\makesavenoteenv{table}
\usepackage{pgfplots}
\usepackage{float}
\usepackage[mathletters]{ucs}
\usepackage[utf8x]{inputenc}
\usepackage{pifont}
\usepackage{multirow}
\newcommand{\cmark}{\ding{51}}%
\newcommand{\xmark}{\ding{55}}%
\usepackage{amsmath}
\usepackage[T1]{fontenc}    
\usepackage{hyperref}       
\usepackage{url}            
\usepackage{booktabs}       
\usepackage{amsfonts}       
\usepackage{microtype}      
\usepackage{todonotes}
\usepackage{graphicx}
\usepackage{subcaption}
\usepackage{wrapfig}
\usepackage{xcolor}
\usepackage{enumitem}
\usepackage{algorithm2e}
\usepackage{supertabular}
\usepackage[mathscr]{euscript}
\usepackage{newtxmath}
\usepackage{euscript}

\mathchardef\mhyphen="2D

\usepackage{listings}
\usepackage{color}

\definecolor{diffgreen}{rgb}{0.4,1.0,0}
\definecolor{diffred}{rgb}{1.0,0.03,0}
\definecolor{gray}{rgb}{0.5,0.5,0.5}
\definecolor{mauve}{rgb}{0.58,0,0.82}
\definecolor{codegreen}{rgb}{0,0.6,0}
\definecolor{codeblue}{rgb}{0,0,0.7}
\definecolor{codegray}{rgb}{0.5,0.5,0.5}
\definecolor{codepurple}{rgb}{0.58,0,0.82}
\definecolor{backcolour}{rgb}{0.95,0.95,0.92}

\newcommand{\RhoSynth}{\textsc{RhoSynth}}
\newtheorem{theorem}{Theorem}
\newtheorem{lemma}[theorem]{Lemma}

\begin{document}

\title{Example-based Synthesis of Static Analysis Rules}         


\author{Pranav Garg}
\affiliation{
  \institution{AWS AI}            
  \streetaddress{410 Terry Avenue North}
  \city{Seattle}
  \state{WA}
  \postcode{98109}
  \country{USA}                    
}
\email{prangarg@amazon.com}          

\author{Srinivasan Sengamedu SHS}
\affiliation{
  \institution{AWS AI}           
  \streetaddress{410 Terry Avenue North}
  \city{Seattle}
  \state{WA}
  \postcode{98109}
  \country{USA}                    
}
\email{sengamed@amazon.com}         


\begin{abstract}
Static Analysis tools have rules for several code quality issues and these rules are created by experts manually. In this paper, we address the problem of automatic synthesis of code quality rules from examples.
We formulate the rule synthesis problem as synthesizing first order logic formulas over graph representations of code. We present a new synthesis algorithm \RhoSynth\ that 
is based on Integer Linear Programming-based graph alignment for identifying code elements of interest to the rule.
We bootstrap \RhoSynth\  by leveraging code changes made by developers as the source of positive and negative examples.
We also address rule refinement in which the rules are incrementally improved with additional user-provided examples. 
We validate \RhoSynth\ by synthesizing more than 30 Java code quality rules.
These rules have been deployed as part of a code review system in a company and their precision exceeds 75\% based on developer feedback collected during live code-reviews.
Through comparisons with recent baselines, we show that current state-of-the-art program synthesis approaches are unable to synthesize most of these rules.

\end{abstract}

\ccsdesc[500]{Software and its engineering~General programming languages}
\ccsdesc[300]{Social and professional topics~History of programming languages}

\keywords{Program synthesis, Programming by example}  

\maketitle

\section{Introduction}

Software is an integral part of today's life and low code quality has  large costs associated with it. For example, the Consortium for Information \& Software Quality notes that ``For the year 2020, the total Cost of Poor Software Quality (CPSQ) in the US is \$2.08 trillion.''\footnote{\url{https://www.it-cisq.org/pdf/CPSQ-2020-report.pdf}}  
Hence automated tools to detect code quality issues are an important part of software development. 
Developers use a variety of tools to improve their software quality, ranging from linters (e.g., SonarLint) to general-purpose code analysis tools (e.g., FindBugs, ErrorProne and Facebook Infer) to specialized tools (e.g., FindSecBugs for code security).  
There are many facets to code quality and often developers want to develop custom analyzers which detect code quality issues that are of specific interest to them.

A common approach for developing custom analyzers or checkers is expressing them in a domain specific language (DSL), such as Semmle's  CodeQL\footnote{\url{https://codeql.github.com/}} or Semgrep\footnote{\url{https://semgrep.dev/}}. 
Code analysis tools such as SonarQube and Semgrep have  thousands of rules that have been manually developed by the community of users or the product owners. Development of these rules is a continuous process:
new rules are added to guide usage of new APIs or frameworks, and existing rules evolve as the APIs evolve. 
Further, expressing code checks in a DSL requires specialized knowledge and extensive testing. 
On the whole, the development of rules is tedious and expensive, 
and the amount of required human intervention limits the scalability of this manual approach.

In this paper, we address the problem of developing code quality rules 
from labeled code examples. 
These examples can be provided by rule authors in the form of conforming and violating examples. 
In most cases, providing these examples is easier than writing the rule itself. In fact, often, these rules are developed following test-driven development (TDD)~\cite{tdd}, where the conforming and violating code examples are used to guide the development of these rules~\cite{pvsstudio}. 
Rules from various tools such as SonarQube and Semgrep, actually, come with such labeled test code examples. 
Additionally, we can leverage \textit{code changes} as the source of labeled examples. 
In a corpus of  code changes, common code quality issues will have multiple code changes that fix those issues.
Similar code changes that are performed by multiple developers across projects can be used to obtain examples to synthesize code quality rules.
For a code change (\texttt{\small code-before}, \texttt{\small code-after}) that fixes a code quality issue, \texttt{\small code-before} is a violating example and \texttt{\small code-after} is a conforming example. 

We propose \RhoSynth, an algorithm for automatically synthesizing high-precision rules from labeled code examples. 
\RhoSynth\ reduces the rule synthesis problem to synthesizing first order logic formulas over Program Dependence Graphs (PDGs). 
Using a  PDG based program representation not only allows our approach to be more robust to syntactic variations in code, but it also enables a succinct expression of semantic information such as data-flow relation and control-dependence. 
Compared to existing approaches that synthesize bug-fixes on Abstract Syntax Trees (ASTs)~\cite{getafix,revisar,refazer,lase}, synthesizing rules over a graph representation comes with its own challenge.
Unlike trees, where a node can be  uniquely identified with the path from the root of the tree to itself, there is no such unique id for nodes in a graph. 
This ambiguity in identifying a node, consequently identifying a subgraph, makes the synthesis problem hard~\cite{haussler}. 
We use Integer Linear Programming (ILP)~\cite{ilp_book,ilp} to solve this problem that lies at the core of rule synthesis. 

We express rules as first order logic formulas comprising an existentially quantified precondition and an existentially quantified postcondition. 
Specifically, rules have the following format: $\exists \vec{x}. pre(\vec{x}) \wedge \neg \big(\bigvee_i \exists \vec{y}. post_i(\vec{x}, \vec{y}) \big)$, where $\vec{x}$ and $\vec{y}$ denotes   a set  of nodes in the PDG. 
The precondition captures the applicability of the rule in violating code examples and the postcondition captures the pattern that \textit{must} be present in conforming code, if the code example satisfies the precondition. 
This is a rich format that can express a wide range of code quality issues. 
We synthesize such rules by first synthesizing a precondition $\exists \vec{x}. pre(\vec{x})$ over violating examples, and then synthesizing the postcondition $\exists \vec{y}. post_i(\vec{x}, \vec{y})$ that accepts conforming examples satisfying $pre(\vec{x})$ for a given valuation $\vec{x}$. Through this decomposition, we simplify the rule synthesis problem to synthesizing existentially quantified precondition and postcondition formulas, which are themselves synthesized using ILP.
Further, the postcondition formula may contain disjunctions. We propose a top-down entropy-based algorithm to partition conforming examples into groups and synthesize a conjunctive postcondition disjunct for each group. 

Recently, there has been lots of work on automated program repair from code changes~\cite{getafix,revisar,refazer,phoenix,lase}. 
While code changes themselves are a good source of examples for synthesizing code quality rules, there may be variations in correct code that are not captured in code changes. 
We propose \textit{rule refinement} to incrementally improve the rule
by providing additional examples corresponding to such code variations.
It turns out, both ILP-based graph alignment and disjunctive postcondition synthesis are instrumental in leveraging "unpaired" code examples for refining the rules.

The main contributions of the paper are as follows.
\begin{enumerate}
\item We formulate the problem of synthesizing rules from labeled code examples as synthesis of logical formulas over graph representations of code.
We present a novel algorithm \RhoSynth\ that is based on ILP based graph alignment, for identifying nodes in the graph that are relevant to the rule being synthesized.

\item We propose \textit{rule refinement} which improves the precision of rules based on a small number of labeled false positive examples.

\item We validate our algorithm by synthesizing more than 30 Java code-quality rules from labeled code examples obtained from code changes in GitHub packages. 
These rules have been deployed as part of a code review system in a company. 
We validate these synthesized rules based on offline evaluation as well as live code review feedback collected over a period of several months. The precision of synthesized rules exceeds 75\% in production. Rule refinement improves the precision of rules by as much as 68\% in some cases.

\item We show that recent baselines can synthesize only 22\% to 61\% of the rules. 
In addition, we show that, compared to ILP-based graph alignment, commonly used tree-differencing approaches do not perform well when aligning unpaired code examples for rule refinement and this results in rules not being synthesized in a majority of cases.

\end{enumerate}

The paper is organized as follows.  Section~\ref{overview:sec} formally defines the rule synthesis problem and provides an outline of the approach. Section~\ref{prog-rule-rep:sec} describes the representation for programs and rules. Section~\ref{ruleSynAlgo:sec} describes the rule synthesis algorithm. 
Section~\ref{sec:implementation} describes the implementation details. 
Section~\ref{expt:sec} describes the experimental results. 
Section~\ref{relwork:sec} presents related work and
Section~\ref{conc:sec} concludes the paper.

\section{\RhoSynth\ Overview} \label{overview:sec}
This section precisely defines the problem and provides an overview of the approach with a running example.

\subsection{Problem Statement}

We are given a set of violating or buggy 
code examples $\mathscr{V} = \{V_1, \cdots, V_m\}$ and a set of 
conforming or non-buggy code examples $\mathscr{C} = \{C_1, \cdots, C_n\}$, for a \textit{single code quality issue}. 
The problem is to synthesize a rule $R$ from a subset of examples such that $R(V_i)$ = \texttt{\small True} and $R(C_j)$ = \texttt{\small False}, for all $i$ and $j$ in the held-out test set. 
We consider code examples at the granularity of a method. 
This offers sufficient code context to precisely capture a wide range of code quality issues~\cite{tufano,defects4j}, while at the same time being simple enough to facilitate efficient rule synthesis.

The corpus of code changes made by developers is a natural source of positive and negative examples.
In such a corpus, common code quality issues will have multiple code changes that fix those issues.
It is possible to obtain examples for \textit{single code quality issues} in an automated manner by clustering code changes~\cite{code-change-clustering,deepcode-pldi18,getafix}. 
Code changes in the input consist of pairs $(B_i, A_i)$ where $B_i$ is the code-before and $A_i$ is the code-after.
To synthesize a rule $R$, we consider \texttt{\small code-before}s as violating examples and \texttt{\small code-after}s as conforming examples. 

A user can also provide additional \textit{conforming examples} corresponding to variations in correct code that  may not be captured by the code changes.

\subsection{\RhoSynth\ Steps}
\label{sec:motivating_example}


Consider the two code changes shown in Figure~\ref{fig:examples}(a)-(b). The code before the change does not handle the case when the cursor accessing the result set of a database query is empty. Without this check, the app might crash when subsequent operations are called on the cursor (e.g., \texttt{\small getString} call on line $8$ in Figure~\ref{fig:examples}(a). In these code changes, the developer adds this check by handling the case when \texttt{\small Cursor.moveToFirst()} returns \texttt{\small False}. 

\begin{figure*}[th]
(a)
\begin{minipage}{.42\textwidth}
\begin{lstlisting}[language = Java,escapechar=\#]
 1	...
 2	Cursor cursor = cr.query(...);
 3#\colorbox{diffred}{\parbox{\textwidth}{-		cursor.moveToFirst();}}#
 4#\colorbox{diffgreen}{\parbox{\textwidth}{+	if ($!$cursor.moveToFirst()) \{ }}# 
 5#\colorbox{diffgreen}{\parbox{\textwidth}{+	\hspace{0.3cm}cursor.close();}}#
 6#\colorbox{diffgreen}{\parbox{\textwidth}{+	\hspace{0.3cm}return null;}}#
 7#\colorbox{diffgreen}{\parbox{\textwidth}{+		\}}}#
 8	final String id = cursor.getString(0);
\end{lstlisting}
\end{minipage}
\hfill
(b)
\begin{minipage}{.47\textwidth}
\begin{lstlisting}[language = Java,escapechar=\#]
 1	...
 2#\colorbox{diffred}{\parbox{\textwidth}{-		mProviderCursor.moveToFirst();}}#
 3#\colorbox{diffgreen}{\parbox{\textwidth}{+	if ($!$mProviderCursor.moveToFirst()) \{ }}# 
 4#\colorbox{diffgreen}{\parbox{\textwidth}{+	\hspace{0.3cm}return;}}#
 5#\colorbox{diffgreen}{\parbox{\textwidth}{+		\}}}#
 6	do {
 7		if(mProviderCursor.getLong(...) == id) {
 8 	... 
\end{lstlisting}
\end{minipage}
\vspace{-0.4cm}
\caption{Examples of input code changes.}
\label{fig:examples}
\end{figure*}

As mentioned earlier, \RhoSynth\ first synthesizes the precondition from buggy code examples and then synthesizes the postcondition from non-buggy code examples.

\begin{figure*}
\begin{minipage}{.8\textwidth}
\resizebox{\textwidth}{!}{
   \includegraphics{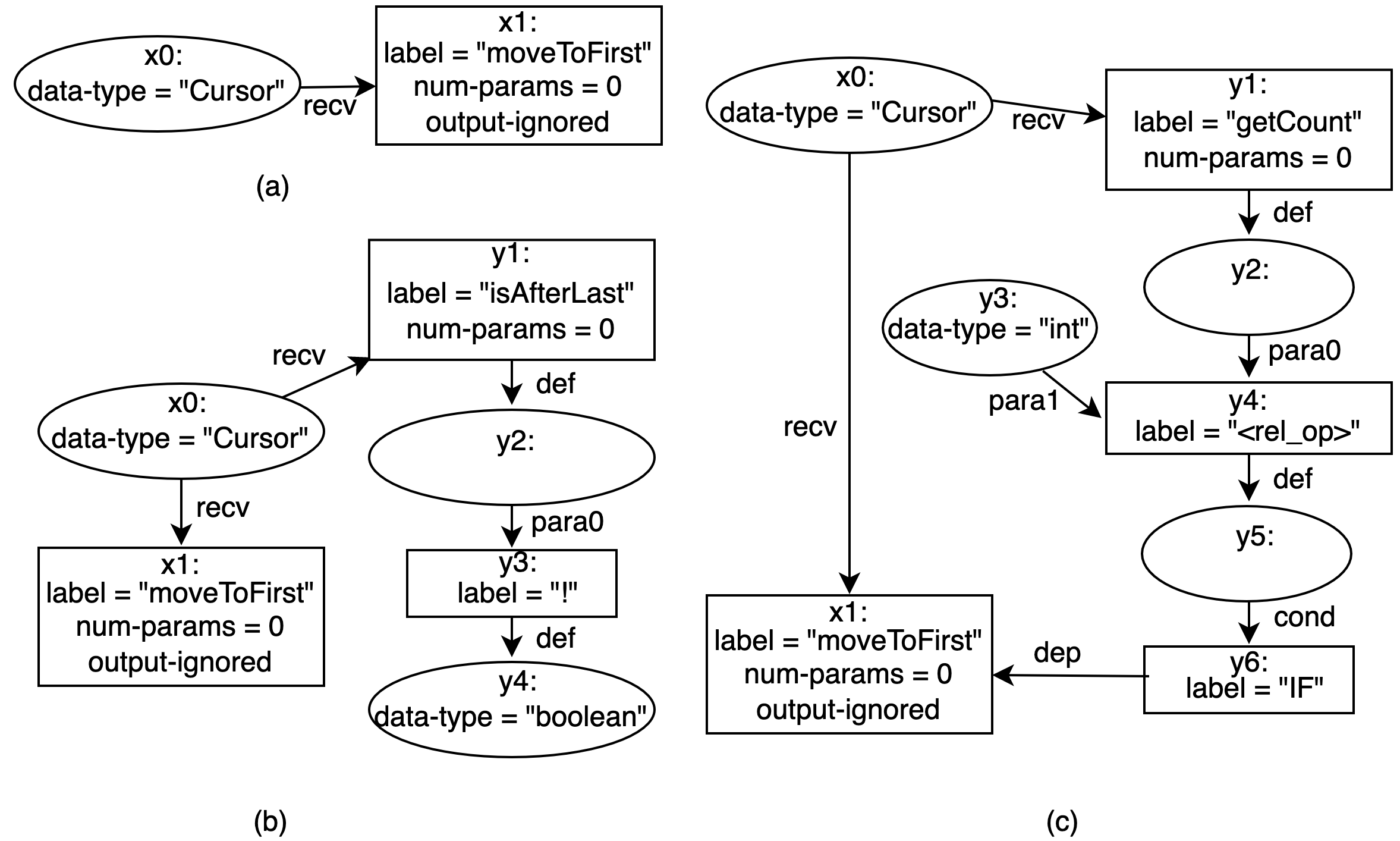}
}
\end{minipage}\\
\begin{minipage}{\textwidth}

\begin{displaymath}
\small
\begin{array}{l}
R = \exists x_0, x_1. \big[ pre(x_0, x_1) ~\wedge \neg \big(\exists y_1, \cdots, y_4. post_1(x_0, x_1, y_1, \cdots, y_4)~~ \vee   \\
 \hspace{4.6cm} \exists y_1, \cdots, y_6. post_2(x_0, x_1, y_1, \cdots, y_6) \big) \big],  \textit{ where}
 \\[8pt]
 
pre(x_0, x_1) := \textit{data-type}(x_0) = \textit{"Cursor"} \wedge \textit{label}(x_1) = \textit{"moveToFirst"} \wedge  \textit{num-para}(x_1) = 0 \wedge \\
    \hspace{0.3cm} \textit{output-ignored}(x_1) \wedge x_0 \xrightarrow{\textit{recv}} x_1\\[8pt]
    
  post_1(x_0, x_1, y_1, \cdots, y_4) := \textit{label}(y_1) = \textit{"isAfterLast"} \wedge  \textit{num-para}(y_1) = 0 \wedge \textit{data-type}(y_2) = \textit{".*"} \wedge\\  
    \hspace{0.3cm} \textit{label}(y_3) = \textit{"!"} \wedge \textit{data-type}(y_4) = \textit{"boolean"} \wedge x_0 \xrightarrow{\textit{recv}} y_1 \wedge y_1 \xrightarrow{\textit{def}} y_2 \wedge y_2 \xrightarrow{\textit{para0}} y_3  \wedge  y_3 \xrightarrow{\textit{def}} y_4 \\[8pt]
   
post_2(x_0, x_1, y_1, \cdots, y_6) := \textit{label}(y_1) = \textit{"getCount"} \wedge  \textit{num-para}(y_1) = 0 \wedge \textit{data-type}(y_2) = \textit{".*"} \wedge\\  
    \hspace{0.3cm} \textit{data-type}(y_3) = \textit{"int"} \wedge \textit{label}(y_4) = \textit{"<rel\_op>"} \wedge \textit{data-type}(y_5) = \textit{".*"} \wedge \textit{data-type}(y_6) = \textit{"IF"} \wedge  \\
	\hspace{0.3cm} x_0 \xrightarrow{\textit{recv}} y_1 \wedge y_1 \xrightarrow{\textit{def}} y_2 \wedge y_2 \xrightarrow{\textit{para0}} y_4  \wedge  y_3 \xrightarrow{\textit{para1}} y_4 \wedge y_4 \xrightarrow{\textit{def}} y_5 \wedge  y_5 \xrightarrow{\textit{cond}} y_6 \wedge  y_6 \xrightarrow{\textit{dep}} x_1  \\
	\hspace{6cm}  \textrm{(d)}
\end{array}
\end{displaymath}
\end{minipage}
\caption{Rule synthesized from code examples in Figure~\ref{fig:examples}: (a) UAPDG for the rule precondition (b) UAPDG for the first disjunct in the rule postcondition (c) UAPDG for the second disjunct in the rule postcondition (d) Overall rule, after refinement, expressed in logic.}
\label{fig:overview}
\end{figure*}

\medskip\noindent\textbf{Rule Precondition Synthesis:}
The precondition synthesis uses only buggy code examples.
We perform a graph alignment on their PDG representations to know the correspondence between nodes in different examples. The graph alignment is framed as an ILP optimization problem.  In our 
\begin{wrapfigure}{h}{0.43\textwidth}
\begin{center}
\scalebox{0.2}{
	\includegraphics{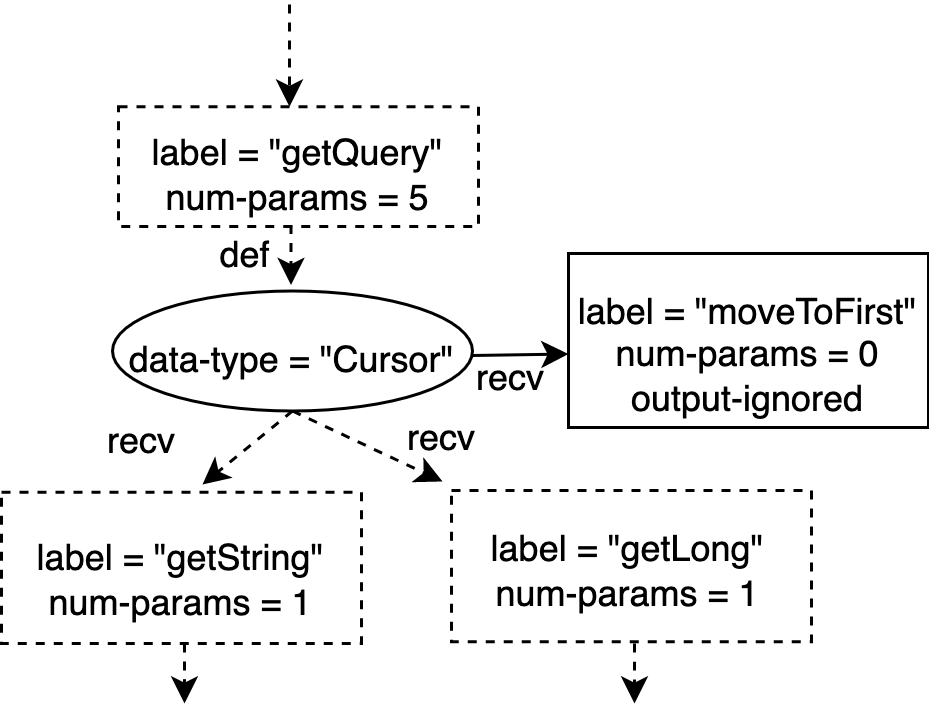}
}
\end{center}
\caption{Unified Annotated PDG (UAPDG) representation that captures \emph{all} code-before's in the input. Dashed lines indicate that the corresponding nodes and edges are present in only a subset of the examples.}
   \label{fig:apdg-precondition}
\end{wrapfigure}
\noindent example, graph alignment determines that the data variables \texttt{\small cursor} in Figure~\ref{fig:examples}(a) and \texttt{\small mProviderCursor} in Figure~\ref{fig:examples}(b) correspond to each other. Similarly, the calls \texttt{\small moveToFirst} correspond. On the other hand, call \texttt{\small getString} in the first example does not have any corresponding node in the second example. 
We use this node correspondence map to construct a Unified Annotated PDG (UAPDG) representation that encapsulates information from all buggy examples. Figure~\ref{fig:apdg-precondition} partially illustrates this UAPDG $\mathscr{A}$. \textit{The solid lines in the figure indicate that the corresponding nodes and edges are present in all buggy examples.} We project $\mathscr{A}$ to these solid nodes and edges and  obtain $\mathscr{A}_c$, which is shown in Figure~\ref{fig:overview}(a). $\mathscr{A}_c$ corresponds to the precondition formula $\exists x_0, x_1. \textit{pre}(x_0, x_1)$ where $\textit{pre}(x_0,x_1)$ is described in Figure~\ref{fig:overview}(d). Besides other checks, the precondition asserts that the output of \texttt{\small moveToFirst} call is ignored, i.e., it is not defined or not used.

\begin{figure}[th]
\begin{minipage}{.48\textwidth}
\begin{lstlisting}[language = Java]
 1	...
 2	Cursor c = sql.query(...);
 3	c.moveToFirst();
 4	while (!c.isAfterLast()) {
 5		Record record = cursorToCognitoRecord(c); 
 6		recordList.add(record); 
 7		c.moveToNext();
 8	}
 9	...
\end{lstlisting}
\subcaption{~}
\end{minipage}
\hfill
\begin{minipage}{.45\textwidth}
\begin{lstlisting}[language = Java,escapechar=!]
 1	...
 2	Cursor c = cr.query(...);
 3	if (c.getCount() == 1) {
 4		c.moveToFirst();
 5		final String key = c.getString(...);
 6		...
 7	} else {
 8		Log.debug("...");
 9		...
10 }
\end{lstlisting}
\subcaption{~}
\end{minipage}
\vspace{-0.4cm}
\caption{Non-buggy code examples used for refining the originally synthesized rule}
\label{fig:refinement_examples}
\end{figure}

\medskip\noindent\textbf{Rule Postcondition Synthesis:} We now find all non-buggy code examples that satisfy the precondition. We use a satisfiability solver for this. If there are such examples, we synthesize a postcondition from them and strengthen the precondition with $\neg \textit{post}$ so that the overall rule rejects the non-buggy examples.
There are no code-after examples in the input that satisfy the precondition, since the value returned by \texttt{\small moveToFirst} is used in all of these examples. Consequently, we synthesize a vacuous postcondition  = \texttt{\small False} and the overall rule is the same as the precondition we synthesized above.

\medskip\noindent\textbf{Rule Refinement:} 
In several cases, the initial set of examples does not capture all code variations. In case of  our current rule, the following checks also check the emptiness of the result set:
\begin{enumerate}
\item \texttt{\small Cursor.getCount() == 0}.
\item \texttt{\small Cursor.isAfterLast()} returns \texttt{\small True}. 
\end{enumerate}

These variations are not part of the initial examples. When we run the synthesized rule on code corpus, we encounter examples such as the ones shown in Figure~\ref{fig:refinement_examples}(a)-(b) that check the cursor using these code variations.  Note these examples are not accompanied by buggy code. We propose \textit{rule refinement} that uses these additional non-buggy examples to improve the rule.
Specifically, we re-synthesize the postcondition by constructing an UAPDG $\mathscr{A}_{\textit{pc}}$, in a way similar to the UAPDG construction in rule precondition synthesis.
However, it turns out that $\mathscr{A}_{\textit{pc}}$ is too general and accepts even the buggy examples. 
We use this as a forcing function to partition the non-buggy examples and synthesize a conjunctive postcondition for each partition.

Figure~\ref{fig:overview}(a)-(c) illustrate the UAPDGs for the synthesized precondition and the two disjuncts in the postcondition. Figure~\ref{fig:overview}(d) provides the exact rule,  expressed as a logical formula.  
Informally, the synthesized $R$ satisfies all code examples that call \texttt{\small Cursor.moveToFirst} such that they do not check the value returned by this call, nor call \texttt{\small Cursor.isAfterLast} or \texttt{\small Cursor.} \texttt{\small getCount}.
When we run this rule again on the code examples, it correctly accepts all the buggy examples and rejects all non-buggy examples, including the additional examples that were used for rule refinement.

\section{Program and Rule Representation} \label{prog-rule-rep:sec}

In this section, we describe the PDG based representation of code (Section~\ref{pdg:sec}), the rule syntax (Section~\ref{ruleRep:sec}) and introduce Unified Annotated PDGs (UAPDGs) for representing rules (Section~\ref{apdg:sec}).

\subsection{Code Representation} \label{pdg:sec}

We represent code examples at a method granularity using PDGs~\cite{pdg}.
PDG is a labeled graph that captures all data and control dependencies in a program. 
Nodes in the PDG are classified as \textit{data nodes} and \textit{action nodes}. Data nodes are, optionally, labeled with the data types and values for literals, and action nodes are labeled with the operations they correspond to, for e.g., method name for method call nodes, etc. 
Edges in the PDG correspond to data-flow and control dependencies and are labeled as
\textit{recv} (connects receiver object to a method call node), \textit{para}$_i$ (connects the $i^{th}$ parameter to the operation), \textit{def} (connects an action node to the data value it defines), \textit{dep} (connects an action node to all nodes that are directly control dependent on it) and \textit{throw} (connects a method call node to a \texttt{\small catch} node indicating exceptional control flow).
See Figure~\ref{fig:pdg} in Appendix~\ref{appendix:pdg} for the PDG of  code-after in Figure~\ref{fig:examples}(a).

\begin{figure}[t!h!]
\begin{displaymath}
\begin{array}{rlll}
\textrm{Rule :} & R &  \stackrel{\textit{def}}{=} &  \exists \vec{x}. \textit{pre}(\vec{x}) \wedge \neg \big(\bigvee\limits_i \exists \vec{y}. \textit{post}_i(\vec{x}, \vec{y})\big), \textit{ where }  \\ 
& & & \hspace{0.3cm} \textit{pre}(\vec{x}) \in \varphi(\vec{x}), \hspace{0.15cm} \textit{ and } \hspace{0.15cm} \textit{post}_i(\vec{x}, \vec{y}) \in \varphi(\vec{x},\vec{y})
 \end{array}
 \end{displaymath}
 \begin{displaymath}
\begin{array}{rl}
\textrm{Conjunctive} & \\
\textrm{Subrule:} & \varphi(\vec{x})   ::=   \bigwedge_i \varphi^i(\vec{x}) \mid  \epsilon(\vec{x}) \mid  \eta(\vec{x}) \mid  \texttt{\small True} \mid \texttt{\small False} \\
\textrm{Edge predicate:} & \epsilon(\vec{x})   ::= ~~~  x_1 \xrightarrow{e} x_2 \\
\textrm{Node predicate:} & \eta(\vec{x})   ::=   ~~~\textit{label(x)} = l \mid \textit{data-type(x)} = R \\
& \mid \textit{data-value(x)} = s  \mid \textit{num-para(x)}  = i \\
&   \mid \textit{declaring-type(x)}  = R\\
&   \mid \textit{trans-control-dep(x)}  \supseteq 2^{Label} \\
& \mid \textit{output-ignored(x)} \mid \neg \textit{output-ignored(x)}
\end{array}
 \end{displaymath}
 \begin{displaymath}
\begin{array}{c}
\vec{x} = \{x_1,\cdots\}, \vec{y} = \{y_1, \cdots\}, x_i, y_i \in \textit{Var}, l \in \textit{Label}, \\
s \in \textit{String}, i \in \textit{Int}, R \textit{ is regular expression}, \\
e \in  \{ \textit{recv, para$_i$, def, dep, throw} \}  
\end{array}
\end{displaymath}
\caption{Syntax of Rules 
}\label{fig:rule_syntax}
\vspace{-0.3cm}
\end{figure}

\subsection{Rule Syntax}	 \label{ruleRep:sec}

In this work, we express rules as quantified first-order logic formulas over PDGs
 (refer to Figure~\ref{fig:rule_syntax} for a detailed syntax of rules). 
A rule is a formula of the form $\exists \vec{x}.{pre}(\vec{x}) \wedge \neg \big( \bigvee \exists \vec{y}. post(\vec{x},\vec{y}) \big)$, where $\vec{x}$ and $\vec{y}$ are a set of quantified variables that range over distinct nodes in a PDG. The precondition $\textit{pre}(\vec{x})$ evaluates to \texttt{\small True} on buggy code and the postcondition $\textit{post}_i(\vec{x}, \vec{y})$ evaluates to \texttt{\small True} on correct code, with appropriate instantiations for $\vec{x}$, $\vec{y}$. Because of the negation before the postcondition, the entire formula evaluates to \texttt{\small True} on buggy code and \texttt{\small False} on correct code. 
Intuitively, the precondition captures code elements of interest in buggy, and possibly correct, code, and the postcondition captures the same in correct code. The elements appear as existentially quantified variables. This is a rich format that can express a wide range
of code quality issues.

Formulas $pre(\vec{x})$ and $post_i(\vec{x},\vec{y})$  are quantifier-free sub-rules comprising a conjunction of atomic edge predicates 
$\epsilon(\vec{x})$ that correspond to edges $x_1 \xrightarrow{e} x_2$ in the PDG, and atomic node predicates $\eta(\vec{x})$.
$\eta({\vec{x}})$ express various properties at PDG nodes including the node label, data-type, data values for literals, number of parameters for method calls (\textit{num-para(x)}), declaring class type for static method calls ($\textit{declaring-type(x)}$), the set of nodes on which $x$ is transitively control dependent (\textit{trans-control-dep(x)}) and, finally, whether a method call's output is/is not ignored (\textit{output-ignored}(x) and its negation)\footnote{Predicate 
\textit{output-ignored}(x) is \texttt{\footnotesize True} for a method call when it does not return a value or the returned value has no users, and \texttt{\footnotesize False} otherwise.}.

Note, we exclude disjunctions in preconditions since a rule with a disjunctive precondition can be expressed as multiple rules without a disjunctive precondition, in the rule syntax. 
Further, a rule with a precondition comprising a negated code pattern (e.g., $\neg x_1 \xrightarrow{e} x_2$) is expressed using a positive pattern in the rule postcondition, and vice-versa.
The rule syntax, while mostly captures first-order logic properties over PDGs, includes some higher-order properties, e.g., transitive control dependence.

\medskip \noindent Checking if a rule satisfies a code example represented as a PDG can be reduced to satisfiability modulo theories (SMT)~\cite{smt}. Since PDGs are finite graphs, this satisfiability check is decidable. 
By mapping nodes in the PDG to bounded integers, this check can be reduced to satisfiability in the Presburger arithmetic.
Moreover, state-of-the-art SMT solvers such as Z3~\cite{z3} can discharge these checks  efficiently.

\subsection{Rule Representation} \label{apdg:sec}

Rules are formulas that accept or reject code examples represented as PDGs. 
From the rule syntax (Figure~\ref{fig:rule_syntax}), a rule is expressed using a precondition and a postcondition.  Both the precondition and the postcondition can be expressed as a collection of quantified formulas $\mathcal{P}_Q = \exists \vec{y}. \varphi(\vec{x}, \vec{y})$, where $\varphi$ is a conjunctive subrule defined over free variables $\vec{x}$ and bound variables $\vec{y}$.
To represent rules, we need a representation for $\mathcal{P}_Q$. 
We arrive at a representation for $\mathcal{P}_Q$, from a PDG representation, in two steps. 
We first introduce Valuated PDGs (VPDGs; also, sometimes,  referred as valuated examples) that are PDGs extended with variables. VPDGs capture a \textit{single} PDG that is accepted by $\mathcal{P}_Q$.  We then introduce Unified Annotated PDGs (UAPDGs) 
that accept \emph{sets} of VPDGs and represent the quantified formula $\mathcal{P}_Q$.

Let $\textit{Var}$ be a set of variables.
A Valuated PDG is a PDG with a subset of its nodes mapped to distinct variables in $\textit{Var}$. Informally, a  Valuated PDG is a PDG with a valuation for variables.

As mentioned above,  a Unified Annotated PDG is defined over a set of free variables $\textit{Var}_f \subseteq \textit{Var}$ and bound variables $\textit{Var}_b \subseteq \textit{Var}$, and represents an existentially quantified conjunctive formula over PDGs of the form $\exists \textit{Var}_b. \varphi(\textit{Var}_f, \textit{Var}_b)$.  Note that
$\varphi$ is a conjunctive subrule that is defined over a set of node predicates ($\eta(\vec{x})$ in Figure~\ref{fig:rule_syntax}). These node predicates are expressed using lattices annotating different nodes in a UAPDG (see Section~\ref{sec:implementation} for details). Formally, 


\begin{definition}[Unified Annotated PDG]
Given a distinct set of free variables $\textit{Var}_f$ and bound variables $\textit{Var}_b$, a $\textit{Unified Annotated PDG} = (N, E, Lat: N \rightarrow (Lattice_1 \times \cdots \times Lattice_k), M_f: N \mapsto \textit{Var}_f, M_b: N \mapsto \textit{Var}_b)$, where $Lat(n)$ is the set of node predicates at $n$ expressed using lattices, and $M_f$ and $M_b$  are maps from a disjoint subset of its nodes to distinct variables in $\textit{Var}_f$ and $\textit{Var}_b$,   respectively. 
\end{definition}

Note that bound variables $\textit{Var}_b$ can be permuted in the map $\textit{M}_b$ without affecting the semantics of a UAPDG. On the other hand, variables $\textit{Var}_f$ cannot be permuted or renamed in the map $\textit{M}_f$, since these variables are free (they are bound to  an outer existential quantifier in the rule $R$).  We call nodes in $\textit{M}_f$, mapped to variables in $\textit{Var}_f$, \textit{frozen} since the variables mapped to these nodes are fixed for a given UAPDG.

For constructing a UAPDG from a set of VPDGs, we need to first \textit{identify corresponding nodes} in VPDGs and then \textit{unify} the node predicates at the corresponding nodes. We identify corresponding nodes through graph alignment using ILP  (see Section~\ref{sec:ilp}) and the unification of node predicates is performed by generalization over the lattice.
We describe the use of UAPDG representation to synthesize rule preconditions and postconditions in Section~\ref{ruleSynAlgo:sec}. 

\medskip\noindent\textit{Notation:} We use italicized font to denote a single example or a PDG or a Valuated PDG (e.g., \textit{C, E, V}) and script font to denote sets of examples or a Unified Annotated PDG (e.g., $\mathscr{A}$, $\mathscr{C}$, $\mathscr{E}$, $\mathscr{V}$). Also, we use superscripts, e.g., $\mathscr{A}^{\vec{x}}$, to denote the set of free variables over which a UAPDG or VPDG is defined ($\mathscr{A}^{\emptyset}$ denotes UAPDG with no free variables).

\begin{example}
Figure~\ref{fig:overview}(b) illustrates a UAPDG defined over free variables $\{x_0, x_1\}$ and bound variables $\{y_1, y_2, y_3, y_4\}$. The label at each node in the figure visualizes the value of maps $\textit{M}_f$, $\textit{M}_b$ and $\textit{Lat}$ at that node.  In this UAPDG,  let $n_1$ be the node with data type "Cursor" and $n_2$ be the node with label "moveToFirst". Then, at node $n_1$, $\textit{M}_f(n_1) = x_0$, $\textit{Lat}(n_1) = (\textit{data-type} = \textit{"Cursor"}, \textit{data-value} = \top, \cdots)$ and map $\textit{M}_b$ is undefined.  Further, we say that nodes $n_1$ and $n_2$ are \textit{frozen} in this UAPDG since they are mapped to free variables $x_0$ and $x_1$ respectively.
\end{example}

\medskip\noindent \textbf{Translating a UAPDG to an existentially quantified formula over graphs:}
A UAPDG naturally translates into an existentially quantified conjunctive formula over Valuated PDGs.  
If  $\textit{Var}_f$ is the set of free variables, 
and $\textit{Var}_b$ is the set of bound variables,  
the translation of the UAPDG is a formula $\exists \textit{Var}_b. \varphi(\textit{Var}_f, \textit{Var}_b)$,
 where $\varphi$ is the conjunction of all satisfying atomic node and edge predicates   expressed 
 over $\textit{Var}_f$ and $\textit{Var}_b$.

\begin{example}
Refer to Figure~\ref{fig:overview}(d) for the quantified rule that corresponds to the UAPDGs in Figure~\ref{fig:overview}(a)-(c). Specifically, the UAPDG in Figure~\ref{fig:overview}(b) is equivalent to the formula: \\
$\exists y_1, \cdots, y_4.  post_1(x_0, x_1, y_1, \cdots, y_4)$.
\end{example}

\medskip\noindent \textbf{Constructing a UAPDG from a Valuated PDG:}
We can construct a UAPDG from a Valuated PDG or a PDG, which can be seen as  a Valuated PDG with an empty node-variable map, in the following manner:
\begin{itemize}
\item Inherit the set of nodes $N$ and the set of edges $E$ from the Valuated PDG.
\item Construct the map $Lat$ by iterating over all nodes $n$ in the Valuated PDG and computing the lattice values for all node predicates at $n$.
\item Initialize $M_f$ with the node-variable map in the Valuated PDG.
\item Construct $M_b$ by mapping all nodes outside the domain of $M_f$ to distinct variables in $\textit{Var}_b$ (these nodes are bound locally to an existential quantifier in the formula that captures the UAPDG). 
\end{itemize}
\noindent Note that a UAPDG obtained by enhancing a PDG (or Valuated PDG), in the above way, corresponds to a formula that accepts the PDG (or Valuated PDG). 
Henceforth, in the text, we use UAPDG to refer to both the graph structure obtained by enhancing PDGs or Valuated PDGs and the existentially quantified formulas they corresponds to, interchangeably.

\section{Rule Synthesis Algorithm} \label{ruleSynAlgo:sec}
The outline of the rule synthesis algorithm is as follows. We first synthesize the precondition based on violating examples. We then synthesize the postcondition based on conforming examples that satisfy the precondition. If the postcondition is not satisfied by any violating example, the synthesis is complete. If a violating example satisfies a postcondition, we follow a hierarchical partitioning approach: pick a conjunct which is satisfied by the violating example, split the underlying partition based on an entropy-based algorithm and recursively synthesize post-conditions for the new partitions. The final postcondition is the disjunction of postconditions of the leaf partitions.

\vspace{0.3cm}
      \begin{minipage}{.5\textwidth}
	\begin{algorithm}[H]
        \scriptsize
        \DontPrintSemicolon
        \SetKwProg{myProc}{Proc}{}{}
        \SetKwFunction{IDthree}{synthesizeRule}
        \SetKwInOut{Input}{input}
        \SetKwInOut{Output}{output}
        \myProc{\IDthree ($\mathscr{V}, \mathscr{C}$)}{
			\nl pre($\vec{x}$), $\mathscr{V}^{\vec{x}}$ := \texttt{getConjunctiveSubRule}($\mathscr{V}^\emptyset$).\; 			

			\nl $\mathscr{C'}^{\vec{x}}$ := $\{ C^{\vec{x}} | C^{\vec{x}} \models$ pre$(\vec{x}), C \in \mathscr{C} \}$.\;
			\nl postcondition := \texttt{synthesizePC}($\mathscr{V}^{\vec{x}}, \mathscr{C'}^{\vec{x}}$).\;
			\nl \textbf{let} postcondition be of the form $\bigvee\limits_i \exists \vec{y}.$ post$_i(\vec{x}, \vec{y}$).\;					\nl R := $\exists \vec{x}. $pre$(\vec{x}) \wedge \neg \bigvee\limits_i \exists \vec{y}.$ post$_i(\vec{x}, \vec{y})$.\;				
                \nl \eIf{all $V \in \mathscr{V}$ satisfy R} {
                        \nl \textbf{return} $R$. \mbox{\ \ \ \ \ \ \ \ \ // R is consistent with $\mathscr{V}$ and $\mathscr{C}$}
                }{
                        \nl \textbf{return} \textsc{Fail}. \mbox{\ \ // R is not consistent with $\mathscr{V}$ and $\mathscr{C}$}
                }
        }
        \caption{\footnotesize Overall rule synthesis algorithm}
		\label{alg:overall}
\end{algorithm}
\end{minipage}\hfill
      \begin{minipage}{.5\textwidth}
	\begin{algorithm}[H]
		\scriptsize
        \DontPrintSemicolon
        \SetKwProg{myProc}{Proc}{}{}
        \SetKwFunction{GetConjunctiveSubRule}{getConjunctiveSubRule}
        \SetKwInOut{Input}{input}
        \SetKwInOut{Output}{output}
        \myProc{\GetConjunctiveSubRule ($\mathscr{E}^{\vec{v_f}}$)}{
	        		\nl $\mathscr{A}^{\vec{v_f}}$ := \texttt{merge}($\mathscr{E}^{\vec{v_f}}$);  $\mathscr{A}^{\vec{v_f}, \vec{v_b}}$ := \texttt{assignVars}($\mathscr{A}^{\vec{v_f}}$, $\vec{v_b}$).\; 
        		\nl $\mathscr{A}^{\vec{v_f}, \vec{v_b}}_c$ := \texttt{project}$_{\mathscr{E}^{\vec{v_f}}}$($\mathscr{A}^{\vec{v_f}, \vec{v_b}}$).\;
			\nl $\phi(\vec{v_f}, \vec{v_b})$ := \texttt{getFormula}($\mathscr{A}^{\vec{v_f}, \vec{v_b}}_c$).\;
			\nl $\mathscr{E}^{\vec{v_f},\vec{v_b}}$ := \{ \texttt{project}$_{E^{\vec{v_f}}}(\mathscr{A}^{\vec{v_f}, \vec{v_b}})$ $\mid$ $E^{\vec{v_f}} \in \mathscr{E}^{\vec{v_f}}$ \}.\;
			\nl \textbf{return} $\phi(\vec{v_f}, \vec{v_b})$, $\mathscr{E}^{\vec{v_f},\vec{v_b}}$.\;
			\BlankLine
	}
	\mbox{\bf Procedure to obtain a conjunctive subrule given a set of}
		\mbox{\bf valuated examples.}
	\end{algorithm}
	\end{minipage}

\vspace{0.3cm}
    \begin{minipage}{.54\textwidth}    
	\begin{algorithm}[H]
	\scriptsize
        \DontPrintSemicolon
        \SetKwProg{myProc}{Proc}{}{}
        \SetKwFunction{IDthree}{synthesizePC}
        \SetKwInOut{Input}{input}
        \SetKwInOut{Output}{output}
        \myProc{\IDthree ($\mathscr{V}^{\vec{x}}, \mathscr{C}^{\vec{x}}$)}{
        		\nl \textbf{if} $\mid \mathscr{C}^{\vec{x}} \mid = 0$ \textbf{then} \textbf{return} \textit{False}.\; 
        		\nl partitionList := [$\mathscr{C}^{\vec{x}}$]; post := \textit{True} .\;
        		\nl \While{True}{
        			\nl partition := partitionList.remove(0).\;
        			\nl \textbf{let} partition be $(\mathscr{C}^{\vec{x}}_1, \cdots, \mathscr{C}^{\vec{x}}_k)$.\;
        			\nl \For{i $\in$ $1 \cdots k$}{
        				\nl post$_i(\vec{x}, \vec{y})$, \_ := \texttt{getConjunctiveSubRule}($\mathscr{C}^{\vec{x}}_i$).\;
        				\nl post := post $\vee$ $\exists \vec{y}.$ post$_i(\vec{x}, \vec{y})$.\;
        			}
        			\nl \textbf{if} all $V^{\vec{x}} \in \mathscr{V}^{\vec{x}}$ satisfy $\neg $ post \textbf{then} \textbf{return} post.\;
					\nl i := index s.t. $V^{\vec{x}} \in \mathscr{V}^{\vec{x}}$ satisfies post$_i$.\;                	
					\nl \textbf{if} $\mid \mathscr{C}^{\vec{x}}_i \mid = 1$ \textbf{then return} \textsc{Fail}.\; 
					\nl \For{($\mathscr{C}^{\vec{x}}_{i, l}$, $\mathscr{C}^{\vec{x}}_{i, r}$) in \texttt{getCandidatePartitions}($\mathscr{C}^{\vec{x}}_i$)}{ 
						\nl partitionList.add(partition $- \mathscr{C}^{\vec{x}}_{i} + \mathscr{C}^{\vec{x}}_{i, l} + \mathscr{C}^{\vec{x}}_{i, r}$)
					} 							
        		}	    
	}    
        \mbox{\bf Procedure to synthesize postcondition from a given set of}
		 \mbox{\bf valuated violating and conforming examples.\label{fig:synthpc}}
	\end{algorithm}
	\end{minipage}      
\hfill
    \begin{minipage}{.52\textwidth}    	
	\begin{algorithm}[H]
		\scriptsize
        \DontPrintSemicolon
        \SetKwProg{myProc}{Proc}{}{}
        \SetKwFunction{IDthree}{generateCandidatePartitions}
        \SetKwInOut{Input}{input}
        \SetKwInOut{Output}{output}
        \myProc{\IDthree ($\mathscr{C}^{\vec{x}}$)}{
			    	\nl $\mathscr{A}^{\vec{x}}$ := \texttt{merge}($\mathscr{C}^{\vec{x}}$).\;
				\nl $\mathscr{A}^{\vec{x}}_c$ := \texttt{project}$_{\mathscr{C}^{\vec{x}}}$($\mathscr{A}^{\vec{x}}$).\;
			    \nl candidateNodes := \{n $\mid$ n is action node, n $\not\in \mathbb{N}$($\mathscr{A}^{\vec{x}}_c$), n $\in \EuScript{N}_{\mathscr{A}^{\vec{x}}}(u)$, u $\in \mathbb{N}$($\mathscr{A}^{\vec{x}}_c$)\}.  \mbox{\ // $\EuScript{N}_{\mathscr{A}}$ is neighbor relation in $\mathscr{A}$} \;
			    \mbox{\ // $\mathbb{N}(\mathscr{A})$ is the set of nodes for $\mathscr{A}$} \;			    			    			    
			    \nl H(n) := \texttt{computeEntropy}(n) \textbf{for } all n $\in$ candidateNodes.\;
			    \nl \For{n $\in$ candidateNodes s.t. H(n) < $\min\limits_{n}$ H(n) + $\delta$}{
			    		\nl partitionList += ($\mathscr{C}^{\vec{x}}_n, \mathscr{C}^{\vec{x}}_{\neg n}$).\;
			    }
			    \nl $\textbf{return}$ partitionList.\;
	}    
	\mbox{\bf Procedure to obtain candidate partitions for a given set of}
		\mbox{\bf valuated conforming examples (see Section~\ref{sec:partition}).\label{fig:getpartitions}}
	\end{algorithm}
\end{minipage}
\vspace{0.3cm}

The overall rule synthesis algorithm is described in Algorithm~\ref{alg:overall}. Inputs to the algorithm is the set of violating $\mathscr{V}$ and conforming examples $\mathscr{C}$. It proceeds in the following steps:

\noindent\textbf{Synthesize precondition from $\mathscr{V}$:}
This is achieved by method \texttt{\small getConjunctiveSubRule()}.
It first calls method \texttt{\small merge} to align examples in $\mathscr{V}$ using ILP (described in detail in Section~\ref{sec:ilp} and~\ref{sec:merge}). This constructs a UAPDG $\mathscr{A}$. Then, it identifies nodes in $\mathscr{A}$ that have a mapping to some node in \textit{each} example. This is performed by method \texttt{\small project} and results in a UAPDG $\mathscr{A}_c$ (subscript $c$ stands for common). Nodes in $\mathscr{A}_c$ are existentially quantified since they correspond to some node in each example in $\mathscr{V}$. 
This resultant UAPDG is translated to the precondition formula $\textit{pre}(\vec{x})$ using the translation described in Section~\ref{apdg:sec}.  \\
\noindent\textbf{Find subset $\mathscr{C'}$ of $\mathscr{C}$ which satisfy the precondition:} 
This is accomplished by querying the SMT solver. If a conforming example satisfies the precondition $\textit{pre}(\vec{x})$, the solver also returns the model comprising the satisfying node assignments for $\vec{x}$. 
We map these nodes to the corresponding variables in $\vec{x}$ to get the set of VPDGs $\mathscr{C}'$.  
Note that the nodes in the VPDGs mapped to variables in $\vec{x}$ are frozen,  since the examples satisfy the precondition through these nodes.\\
\noindent\textbf{Synthesize postcondition from $\mathscr{C'}$:} 
This is achieved by method \texttt{\small synthesizePC()}. Besides $\mathscr{C'}$, we also pass to this method violating examples $\mathscr{V}$ with a node assignment that satisfy the precondition. Initially, all the conforming examples in the input constitute a single partition. For these examples, we use the ILP graph alignment to align unfrozen nodes and synthesize a postcondition (using method \texttt{\small getConjunctiveSubRule} as before). 
If the postcondition is unsatisfiable on all frozen violating examples, we return the synthesized postcondition. Otherwise, it implies that the postcondition synthesized from all examples in the partition is too general to reject violating examples, and we partition the set of conforming examples further. The partitioning algorithm is described in Section~\ref{sec:partition}.

\begin{example}
Consider postcondition synthesis for the rule described in Section~\ref{sec:motivating_example}.  
Calling \texttt{\small getConjunctiveSubrule} on all conforming examples satisfying the precondition returns a UAPDG $\mathscr{A}$ that is same as the synthesized precondition, visualized in Figure~\ref{fig:overview}(a).  Violating examples such as \texttt{\small code-before}s in Figure~\ref{fig:examples} satisfy $\mathscr{A}$. Therefore,  confoming examples are partitioned and a separate postcondition is synthesized for each partition (UAPDGs in Figure~\ref{fig:overview}(b-c)). 
\end{example}

We next describe the different components of the rule synthesis algorithm in greater detail.

\subsection{Maximal Graph Alignment using ILP} 
\label{sec:ilp}

Graph alignment amongst the violating examples or conforming examples is one of the key steps in synthesizing the rule precondition or postcondition respectively. Since both the  precondition and postcondition formulas are existentially quantified, the synthesis problem is NP-hard~\cite{haussler}. Its hardness stems from different choices available for mapping existential variables to nodes in the example graphs.
Alternatively, the synthesis problem can be seen as a search over graph alignments such that aligned nodes in different examples are mapped to the same existential variable.
Note that different graph alignments will result in different synthesized formulas. 
Given multiple graphs, we iteratively align two graphs at a time\footnote{This is an instance of a “multi-graph matching” problem. This requires optimization techniques beyond ILP, for e.g., alternating optimization~\cite{multi_graph_matching}. We did not explore these solutions since pairwise graph alignment worked well for our application.}.
  We frame the problem of choosing a desirable graph alignment for a pair of example graphs as an ILP optimization problem. Since we want to synthesize formulas that precisely capture the code examples, 
we choose alignments that maximize the number of aligned nodes and edges.

The ILP objective to maximize graph alignment can also be viewed as minimizing the graph edit distance between code examples. Consequently, our reduction to ILP follows the binary linear programming formulation to compute the exact graph edit distance between two graphs~\cite{ilp}. The node and edge substitutions are considered cost $0$ and node/edge additions and deletions are cost $1$ operations.
We include a detailed reduction to the ILP optimization problem in Appendix~\ref{app:ilp}. Below,  we describe some constraints on the ILP optimization that are specific to our application:

\begin{itemize}
\item We align action nodes only if they have the same label. So, a method call \texttt{\small foo} in one example cannot align with method \texttt{\small bar} in another example. 
However, data nodes that are,optionally, labeled with their types do not have this restriction. This allows us to align data nodes even when their types do not resolve, or when their types are related in the class hierarchy, for e.g., $\texttt{\small InputStream}$ and $\texttt{\small FileInputStream}$.

\item Two data nodes align only if they have at least one aligned incoming or outgoing edge. This restricts alignment of data nodes to only occurrences when they are defined or used by aligned action nodes.

\item When synthesizing the postcondition, nodes frozen to the same variables in $\vec{x}$ are constrained to be aligned.

\end{itemize}

When synthesizing rules from code changes, we use the ILP formulation described in this section to perform a fine-grained graph differencing over PDGs in a code change (similar to the GumTree tree differencing algorithm~\cite{gumtree}).  Graph differencing labels nodes in the PDGs with change tags: \textit{unchanged}, \textit{deleted} or \textit{added}, depending upon whether the node is present in both code-before and code-after, in only code-before, and in only code-after. Now, while aligning the violating code-before's for synthesizing the precondition, we require that the node alignment respect these change tags, i.e., nodes are aligned only if their change tags are the same.
This constraint helps to focus the synthesized precondition formula on the changed code.
This constraint is removed when synthesizing the  postcondition. This is because even variables $\vec{x}$ frozen in the conforming examples, by design, may not respect these change tags.

\subsection{Synthesizing Conjunctive Subrules}
\label{sec:merge}

In this section, we describe the procedure \texttt{\small getConjunctiveSubRule} to synthesize a conjunctive subrule from a given set of valuated examples.
It has two main steps. First, UAPDGs that correspond to input valuated examples are iteratively \texttt{\small merge}d pairwise to obtain a UAPDG $\mathscr{A}$.
Second, we \texttt{\small project} $\mathscr{A}$ to nodes \textit{present} in all input examples to obtain UAPDG $\mathscr{A}_c$.
Given UAPDGs $\mathscr{A}_1$ and $\mathscr{A}_2$, $\mathscr{A}_c$ obtained by calling \texttt{\small merge} followed by a \texttt{\small project} over-approximates $\mathscr{A}_1 \vee \mathscr{A}_2$. 
This property ensures that $\mathscr{A}_c$ obtained after merging a set of input valuated examples is saltisfied by all of them. We first describe the \texttt{\small merge} operation, followed by \texttt{\small project}.

Let $\mathscr{A}_1 = (N_1, E_1, Lat_1, M_{f1}, M_{b1})$ and $\mathscr{A}_2 = (N_2, E_2, Lat_2$, $M_{f2}, M_{b2})$ be two UAPDGs over the same set of free variables $\textit{Var}_f$.
Let $NM \subseteq N_1 \times N_2$ and $EM \subseteq E_1 \times E_2$ be the node and edge mappings that are obtained from the graph alignment step.
We first extend these mappings to relations $\textit{NM}^{\epsilon} \subseteq N_1 \cup \{\epsilon\} \times N_2 \cup \{\epsilon\}$ and $\textit{EM}^{\epsilon} \subseteq E_1 \cup \{\epsilon\} \times E_2 \cup \{\epsilon\}$ such that 
$\textit{NM}^{\epsilon} = \textit{NM} \cup \{(n_1, \epsilon) \mid (n_1,\_) \not\in NM\} \cup \{(\epsilon, n_2) \mid (\_, n_2) \not\in NM\}$, and
 $\textit{EM}^{\epsilon} = \textit{EM} \cup \{(e_1, \epsilon) \mid (e_1, \_) \not\in EM\} \cup \{(\epsilon, e_2) \mid (\_,e_2) \not\in EM\}$.
Then, $\texttt{merge}_{(NM^\epsilon, EM^\epsilon)}(\mathscr{A}_1, \mathscr{A}_2)$ returns a UAPDG $\mathscr{A} = (N, E, Lat, M_f, M_b)$ constructed as follows:
\begin{itemize}
\item $N = \{(n_1, n_2) |  (n_1, n_2) \in NM^{\epsilon}\}$

\item We first compute $E' = \{(n_1, n_2) \xrightarrow{(e_1, e_2)} (n_1', n_2') |  (e_1, e_2) \in EM^{\epsilon},  e_i \neq \epsilon \Leftrightarrow n_i \xrightarrow{e_i} n_i' \}$. Since $EM$ is an edge mapping obtained from the graph alignment step, it follows that $e_1 \neq \epsilon \wedge e_2 \neq \epsilon \Rightarrow \textit{label}(e_1) = \textit{label}(e_2)$. Further, from definition of $EM^{\epsilon}$, it follows that $e_1 \neq \epsilon \vee e_2 \neq \epsilon$. 
Once $E'$ is computed, $E = \{ (n_1, n_2) \xrightarrow{e} (n_1', n_2') | (n_1, n_2) \xrightarrow{(e_1, e_2)} (n_1', n_2') \in E', e = \{ \textit{label}(e_1), \textit{label}(e_2) \} \textbackslash \{\epsilon\} \}$.

\item $Lat(n_1, n_2) = Lat_1(n_1) \sqcup Lat_2(n_2)$ where the join is applied point-wise to each node predicate  at $n_1$ and $n_2$ (we assume that $Lat(\epsilon) = \bot$)
\item $M_f(n_1, n_2) = x$, if $M_{f1}(n_1) = M_{f2}(n_2) = x \in \textit{Var}_f$
\item 
Nodes not mapped to free variables are mapped to distinct fresh variables-- if $(n_1, n_2) \not\in \textit{domain}(M_f)$, then $M_b(n_1, n_2) = x_i$ for a distinct variable $x_i \not\in \textit{Var}_b$.
\end{itemize}

Method \texttt{\small project} with respect to $\mathscr{A}_1$ and $\mathscr{A}_2$ projects the merged UAPDG $\mathscr{A}$ to $\mathscr{A}_c$ which has nodes (resp. edges) that map to nodes (resp. edges) in both $\mathscr{A}_1$ and $\mathscr{A}_2$, i.e., nodes in $\mathscr{A}_c$ are of the form $(n_1, n_2)$ where $n_i \in N_i$ and edges are of the form $(e_1, e_2)$ where $e_i \in E_i$.

\begin{theorem}\label{thm:merge_soundness}
For any node mapping $NM \subseteq N_1 \times N_2$ and edge mapping $EM \subseteq E_1 \times E_2$ that is obtained from aligning $\mathscr{A}_1$ and $\mathscr{A}_2$, $\mathscr{A}_1 \vee \mathscr{A}_2 \Rightarrow \texttt{project}_{\{\mathscr{A}_1, \mathscr{A}_2\}}(\texttt{merge}_{(NM^\epsilon, EM^\epsilon)}(\mathscr{A}_1, \mathscr{A}_2))$.
\end{theorem}
The proof outline for the above theorem is presented in Appendix~\ref{app:proof1}. 
Besides, returning the formula for $\mathscr{A}_c$, method \texttt{\small getConjunctiveSubrule}   
also returns the set of valuated examples labeled with bound variables used for constructing $\mathscr{A}$. This is achieved by \texttt{\small project}ing $\mathscr{A}$ to nodes (resp. edges) that map to nodes (resp. edges) in each example. 
Note, to synthesize conjunctive subrules from a single example (possible in the case of postcondition synthesis), we heuristically include in the subrule all nodes at distance $d = 1$ from nodes bound to free variables $\vec{x}$.  

\subsection{Partitioning Conforming Examples}
\label{sec:partition}

In this section, we describe the method \texttt{\small generateCandidatePartitions}. 
This method returns a list of low entropy partitions of the input examples $\mathscr{C}$,  based on the clustering algorithm in~\cite{entropy_clustering}.
Let $N$ be the set of nodes in $\mathscr{A}$ and $N_c$ be the set of nodes in $\mathscr{A}_c$, where $\mathscr{A}$ and $\mathscr{A}_c$ are the UAPDGs obtained after calling \texttt{\small merge} and then calling \texttt{\small project} on examples in $\mathscr{C}$ respectively. 
For $n \in N$, let $\mathscr{C}_n = \{C_i\} \subseteq \mathscr{C}$ be the set of examples such that $n$ maps to a node in $C_i$ 
 and let $\mathscr{C}_{\neg n} = \mathscr{C} \backslash \mathscr{C}_n$. Entropy of a partition with respect to $n$ is:\\
$H(\mathscr{C}, \mathscr{C}_n, \mathscr{C}_{\neg n}) = \frac{\mid \mathscr{C}_n \mid}{\mid \mathscr{C} \mid}. H(\mathscr{C}_n) + \frac{\mid \mathscr{C}_{\neg n} \mid}{\mid \mathscr{C} \mid}. H(\mathscr{C}_{\neg n})$, \\
$H(\mathscr{C}) = \sum\limits_{n'} H_{n'}(\mathscr{C})$, where 
$H_{n'}(\mathscr{C}) = -p. log p - (1 - p). log (1 - p)$,\\
 where $p = \mid \mathscr{C}_{n'} \mid / \mid \mathscr{C} \mid$.
Nodes in $N_c$ by definition are mapped to nodes in all input examples. 
Since code patterns in a rule are often localized, we consider partitions with respect to nodes in $N$ that neighbor nodes in $N_c$. Further, we return the set of all partitions whose entropy is within a $\delta$ margin of the smallest entropy partition.
Each of these partitions is explored in a BFS manner in method \texttt{\small synthesizePC}. 
The process stops at the first partition that synthesizes a postcondition  that is unsatisfied by all violating examples.

\subsection{Discussion about \RhoSynth}

\medskip\noindent\textit{Precision and Generalization:}
\RhoSynth\ prefers precision over recall.  For this reason,  \RhoSynth\ biases preconditions to more specific formulas using an ILP-based maximal graph alignment, even if it may cause some amount of overfitting.  In practice,  we minimize overfitting by using diverse examples for synthesizing a rule, for instance,  examples that are obtained from code changes that belong to different packages (refer to Appendix~\ref{input:sec} for details). 
Furthermore,  to prevent overfitting, \RhoSynth\ biases postconditions to fewer disjuncts.  This is accomplished by partitioning a group of correct examples only when the most-specific conjunctive postcondition synthesized from all correct examples satisfies a violating example.  
In this case,  partitioning the correct examples and synthesizing a postcondition disjunct for each partition becomes necessary.
In addition,  \RhoSynth\ uses lattices to express all node predicates.  This helps generalization when merging groups of conforming and violating examples to synthesize the rule.

\medskip\noindent\textit{Soundness:}
We next show that \RhoSynth\ is sound, i.e.,  given a set of violating and conforming examples, if \RhoSynth\ synthesizes a rule $R$ then $R$ satisfies all violating examples and does not satisfy any conforming example.
Formally, 

\begin{theorem}[Soundness of Rule Synthesis:]\label{thm:overall_soundness}
Given a set of violating examples $\mathscr{V} = \{V_1, \cdots, V_m\}$ and conforming  examples $\mathscr{C} = \{C_1, \cdots, C_n\}$, if the algorithm \texttt{\small synthesizeRule} successfully returns a rule $R$, then $V_i \models R$ and $C_i \not\models R$.
\end{theorem}

We present a proof outline in Appendix~\ref{app:proof2}. 
It is easy to argue Algorithm~\ref{alg:overall}'s soundness with respect to violating examples. To argue soundness with respect to conforming examples, we 
rely on the soundness of the \texttt{\small merge} algorithm (theorem~\ref{thm:merge_soundness}). Using Theorem~\ref{thm:merge_soundness}, we argue that 
all valuated conforming examples satisfy the synthesized postcondition. Since the rule negates the synthesized postcondition, none of them satisfy the synthesized rule.

\medskip\noindent\textit{Completeness:}
\RhoSynth\ does not provide completeness guarantees. 
This is a design choice made purely based on practical experiments.
Incomleteness may arise from incorrect graph alignments or incorrect partitioning of conforming examples when synthesizing rule postconditions.  However, in our experiments, we do not encounter cases where either of these lead to rule synthesis failures or imprecise solutions.

\section{Implementation Details}
\label{sec:implementation}

\noindent\textit{PDG representation:}
We use \RhoSynth\ to synthesize rules for Java. Our implementation uses MUDetect~\cite{mudetect} for representing Java source code as PDGs.  MUDetect uses a static single assignment format. Further, we transform these PDGs using the following two program transformations to obtain program representations that are more conducive to rule synthesis:
\begin{enumerate}
\item We abstract the label of relational operators such as $<, \leq, =, \neq$, etc. to a common label \texttt{\small rel\_op}.  This transformation is useful when synthesizing rules from examples that perform a similar check on a data value using different relational operators, for e.g., \texttt{\small error < 0} and \texttt{\small error == -1}.  This  transformation helps to express rules over such examples using a conjunctive formula over the \texttt{\small rel\_op} label,  instead of a disjunction over different relational operators.

\item If a PDG has multiple calls to the same getter method on the same receiver,  we transform it into a PDG that has a single getter call and the value returned by the getter is directly used at other call sites.  
So, we transform the code snippet
$\texttt{\small foo(url.getPort()); ... } \texttt{\small if (url.getPort()}\\ \texttt{\small == 0) \{ ... \} }$ 
to the PDG representation of the code snippet 
$\texttt{\small v = url.getPort();} \texttt{\small foo(v);} \\ \texttt{\small  ...  if (v == 0) \{ ... \} }$. Since \RhoSynth, typically, does not have access to method declarations of called methods,  we use heuristics on the name of the method to identify getter calls.  
\end{enumerate}

\noindent\textit{Lattices:} 
To enable unification of node predicates at corresponding
nodes in different PDGs, we lift all the node predicates to lattices in the following manner: 
\begin{enumerate}
\item Predicate \textit{data-type(x)} and \textit{declaring-type(x)} are lifted to a lattice that captures the longest common prefix and longest common suffix of the string representations of data types. 
Often,  a Java class name shares a common prefix or a suffix with name of its superclass or the interfaces it extends. This lattice is carefully designed to unify such related data types.  
If nodes $n_1$  and $n_2$ in different examples align to form node $n$ where \textit{data-type}($n_1$) = \textit{“BufferedInputStream”} and \textit{data-type}($n_2$) = \textit{“FileInputStream”},  this lattice helps generalize \textit{data-type}(n) = \textit{“*InputStream”}.

\item Predicates \textit{label(x)}, \textit{data-value(x)}, \textit{num-para(x)},
\textit{output-ignored(x)} are lifted to constant lattices.  
As an example,  if \textit{num-para}$(n_1) = 1$  and \textit{num-para}$(n_2) = 2$ and $n_1$ and $n_2$ align to form node $n$,  then \textit{num-para}$(n) = \top$.

\item
Predicate \textit{trans-control-dep(x)} is lifted to a power-set lattice
where set intersection is the join operator.  As an example,  if \textit{trans-control-dep}$(n_1) = \{ \texttt{\small If}, \texttt{\small Catch} \}$  and \textit{trans-control-dep}$(n_2) = \{ \texttt{\small If}\}$ and $n_1$ and $n_2$ align to form node $n$,  then \textit{trans-control-dep}$(n) = \{ \texttt{\small If} \}$.

\end{enumerate}

\noindent\textit{Solvers:}
We use the \textsc{Xpress} ILP solver for graph alignment and \textsc{Z3}~\cite{z3} for satisfiability solving.
For running a rule on a method,  we check if the rule is satisfied by the PDG representation of the method.  
We have implemented a  few algorithmic simplifications that are used to simplify the satisfiability query, 
 for e.g., we can determine that a rule, which asserts existence of a method call \texttt{\small foo},  will not be  satisfied  by a method that does not call \texttt{\small foo}. Due to such simplifications, >99\% of the satisfiability queries can be discharged without calling Z3.

\section{Evaluation} \label{expt:sec}

To understand the effectiveness of \RhoSynth, we address the following research questions.

\begin{enumerate}
\item[\textbf{RQ1:}] How effective is \RhoSynth\ at synthesizing precise rules?

\textit{Section~\ref{sec:rq1} shows that the precision of rules exceeds 75\% in production.}

\item[\textbf{RQ2:}] Are the synthesized rules ``interesting''?

\textit{The rules synthesized cover diverse categories and are often discussed on various forums as shown in Section~\ref{sec:rq1.1}.}

\item[\textbf{RQ3:}] What is the effectiveness of iterative rule refinement in improving the precision?

\textit{As we show in Section~\ref{sec:rq2}, rule refinement improves the precision of rules, by as much as 68\% in some cases.}

\item[\textbf{RQ4:}] How does \RhoSynth\ compare against state-of-the-art program synthesis approaches?
The baselines used are ProSynth~\cite{provenance} and anti-unification based synthesis over ASTs.

\textit{The baselines often fail to synthesize rules.} (Section~\ref{sec:rq3})

\item[\textbf{RQ5:}] How does ILP-based graph alignment compare against state-of-the-art code differencing algorithms? We use the GumTree~\cite{gumtree} algorithm as a baseline. 

\textit{Baseline algorithms do not perform well when aligning unpaired code examples for rule synthesis.} (Section~\ref{sec:rq4})

\end{enumerate}

\subsection{Experimental Methodology and Setup}
\label{sec:methodology}

\medskip\noindent\textit{Code Change Examples:}
We synthesize rules for 31 groups of Java code changes. See Appendix~\ref{input:sec} for a description on the source of the dataset. 
We provide GitHub links to all the code changes along with the synthesized rules as supplementary material.

\medskip\noindent\textit{Iterative Rule Refinement:} 
We run the synthesized rules on code corpus. 
We identify rules for iterative refinement based on the detections generated by them.
For each rule, we first label $5$ detections. If any of these detection is labeled as a false positive (FP), we shortlist the rule for refinement. For such rules, we label up to $10$ additional detections. With every false positive, we resynthesize the rule using the original set of code changes and all the FPs encountered till then. We stop when adding the $N^{\textit{th}}$ FP does not change the synthesized rule, i.e.,  rule synthesized after the first $N$ FPs is equivalent to the rule synthesized after $N-1$ FPs.

\medskip\noindent\textit{Production Deployment:}
We have deployed all these rules in an industrial setting, with our internal code review (CR) system. These rules are run on the source code at the time of each code review. Detections generated by these rules on the code diff are provided as code review comments. Code authors can label these comments as part of code review workflow.

\medskip\noindent\textit{Evaluation:}
We perform two types of evaluation: \textit{offline evaluation} by software developers on sampled recommendations and ratings and textual feedback obtained from \textit{live code reviews} from code authors.

In both cases, the users label them as "Useful", "Not Useful" or "Not Sure". In live code review, developers also provide textual feedback.
The offline evaluation involves $10$ expert developers and does not involve the paper authors.

\medskip\noindent\textit{Metrics:}
For both the evaluations, we report $\textit{precision } = \textit{ \# Useful } / \textit{ \# Total labels}$, computed over all labeled detections.

\medskip\noindent\textit{Experimental Setup:}
We conducted all our experiments on a Mac OSX laptop with 2.4GHz Intel processor and 16Gb memory. Experiments with ProSynth~\cite{provenance} were run in a Docker container with 8Gb memory launched from a Docker image  shared by the authors.

\subsection{Precision of \RhoSynth}
\label{sec:rq1}

In our offline evaluation, $91\%$ detections ($107$ out of $117$) were labeled "Useful".
We have also received feedback from developers on detections generated by these rules during live code reviews. Over a period of $5$ months, detections from $25$ out of the $31$ rules received developer feedback.  $75.8\%$ of these labeled detections ($273$ out of $360$) were categorized as "Useful" by the developers\footnote{Most rules that did not receive developer feedback during live code reviews involved detecting code context with deprecated or legacy APIs. These rules trigger rarely during a code review and are thus more appropriate for a code scanning tool. All these rules were validated during offline evaluation.}. 
Some of the textual feedback from live code reviews are: \lstinline{"Interesting, that is helpful"}, \lstinline{"Will add this check."}, and \lstinline!"we have a separate task to look into the custom polling solution https://XX"!.
A high developer acceptance during code reviews shows that \RhoSynth\ is capable of synthesizing code quality rules which are effective in the real-world.

Table~\ref{fig:rule_synthesis_stats} provides more information about these synthesized rules.
The  synthesis time for these  rules range from $30ms$ to $13s$, with an average  of $1.5s$. 
 Most of these rules are synthesized from few code changes. 
Rules that only consist of a precondition correspond to a code anti-pattern. 
$9$ rules consist of both a precondition and a postcondition.
Of these, $4$ rules have a disjunctive postcondition.  These disjuncts correspond to different ways of writing code that is correct with respect to the rule.  One example of such a rule is \texttt{\small check-movetofirst}, described in Section~\ref{sec:motivating_example}.  Another example is \texttt{\small executor-graceful-shutdown} that intuitively checks if \texttt{\small ExecutorService.shutdownNow()} call is accompanied by a graceful wait implemented by calling \texttt{\small ExecutorService.awaitTermination()} or \texttt{\small ExecutorService.invokeAll()}. 
Besides the supplementary material, we include visualizations for few synthesized rules  in Appendix~\ref{app:apdg}.

\subsection{Examples of \RhoSynth\ rules}
\label{sec:rq1.1}

We describe all the rules synthesized by \RhoSynth\ in Table~\ref{fig:bp_description}. A large number of these rules are supported by code documentation  or discussions on online  forums. They cover a 
wide range of code quality issues and recommendation categories:
\begin{itemize}
\item performance: e.g., use-parcelable, use-guava-hashmap
\item concurrency: e.g., conc-hashmap-put, countdown-latch-await
\item use of deprecated or legacy APIs: e.g., deprecated-base64, upgrade-enumerator
\item bugs: e.g., check-movetofirst, start-activity
\item code modernization: e.g., view-binding, layout-inflater
\item code simplification: e.g., deseriaize-json-array, use-collectors-joining
\item debuggability: e.g.,  countdownlatch-await, exception-invoke
\end{itemize}

\vspace{2cm}

\begin{center}
\scriptsize
\begin{tabularx}{\textwidth}{clll}

\multicolumn{1}{l}{\textbf{S. No.}} & \textbf{Rule name}                         & \textbf{API} & \textbf{Description of the synthesized rules}                                                                                                                                                                                                                                                                                                                                                                                                                                                                                                                                                                                                    \\ \hline

1                                   & 
\begin{tabular}[c]{@{}l@{}} check-\\actionbar        \end{tabular} &
\begin{tabular}[c]{@{}l@{}}Android \\ getSupport-\\ActionBar\end{tabular}                                                                                    & 
\begin{tabular}[c]{@{}l@{}}
The method `getSupportActionBar` returns `null` if the Android activity does not have  \\ an action bar.  One must null-check the 
value returned by `getSupportActionBar`,  if the  \\action bar is not explicitly set by a `setSupportActionBar` call.
\end{tabular}                                                                                                                                      \\\hline

2                                 & 
\begin{tabular}[c]{@{}l@{}}check-\\await-\\ termination\end{tabular}      & 
\begin{tabular}[c]{@{}l@{}}Executor-\\Service.await-\\Termination\end{tabular}                                                                          & 
\begin{tabular}[c]{@{}l@{}}
One must check the return value of `awaitTermination()` to determine if the operation  \\  timed out while waiting for other threads 
to stop execution,  following a  shutdown  \\ request. Alternatively,  one can check the same by calling`ExecutorService.isTerminated`.
\end{tabular}                                                                                                                                                           \\\hline

3                                   & 
\begin{tabular}[c]{@{}l@{}} check-\\create-\\newfile   \end{tabular} &
\begin{tabular}[c]{@{}l@{}}File.\\create-\\NewFile\end{tabular}                                                                                                           &
 \begin{tabular}[c]{@{}l@{}}
`createNewFile` returns False if a file already exists. One must check for `File.exists` or \\  check the value returned by `createNewFile`.
 Without this check,   one might silently  \\ overwrite an existing file leading to a data loss.
\end{tabular}                                                                                                                                                         \\\hline

4                                   & 
\begin{tabular}[c]{@{}l@{}} check-\\file-\\rename     \end{tabular} &
\begin{tabular}[c]{@{}l@{}}FileSystem.\\rename \end{tabular}  & 
\begin{tabular}[c]{@{}l@{}}
One must check if the Hdfs `rename` operation succeeded and handle failures otherwise.\\  This check can be performed  by checking
 the value returned by the `rename`  call.\\   Silent failures can lead to errors that are harder to debug.
\end{tabular}                                                                                                                                                                     \\\hline

5                                  & 
\begin{tabular}[c]{@{}l@{}} check-\\inputstream-\\skip    \end{tabular} &
\begin{tabular}[c]{@{}l@{}}Input-\\Stream.\\skip \end{tabular}                                                                                                           &                                                                                  
\begin{tabular}[c]{@{}l@{}}
`InputStream.skip` returns the number of bytes skipped.  One must check the value  \\ returned by the `skip` call to handle the case 
when  fewer than the expected number of  \\ bytes are skipped.
\end{tabular}                                                                                                                                                                               \\\hline

6                                 & 
\begin{tabular}[c]{@{}l@{}} check-\\mkdirs  \end{tabular} &
\begin{tabular}[c]{@{}l@{}}  File.\\mkdirs \end{tabular}        &
\begin{tabular}[c]{@{}l@{}}
One must check if the `mkdirs` operation succeeded and handle failures otherwise.  This \\check can be performed by checking the 
value returned by the `mkdirs` call.  Silent  \\  failures can lead to errors that are harder to debug.
\end{tabular}                                                                                                                                                                                  \\\hline

7                                   & 
\begin{tabular}[c]{@{}l@{}} check-\\movetofirst      \end{tabular} &
\begin{tabular}[c]{@{}l@{}}  Cursor.\\moveToFirst             \end{tabular}        &                                                                                                                           
\begin{tabular}[c]{@{}l@{}}
One must check if the result set returned by a database query is empty. This can be  \\ performed by checking the value returned by 
`Cursor.moveToFirst` or `Cursor.isAfterLast`, \\  or checking if `Cursor.getCount() > 0`. Without this check,  the app can crash  \\ if 
subsequent operations are called on the cursor.\end{tabular}                                                                                                                                                                              \\\hline

8                                  & 
\begin{tabular}[c]{@{}l@{}} check-\\resultset-\\next      \end{tabular} &
\begin{tabular}[c]{@{}l@{}}  ResultSet.\\next \end{tabular}        &
\begin{tabular}[c]{@{}l@{}}One must check if `ResultSet.next()` returns False.  If it returns False, it implies that \\ the cursor is positioned after the last row and
 any subsequent calls to `next()` will  \\ throw an exception.\end{tabular}                                                                                                                                            \\\hline

9                                  & 
\begin{tabular}[c]{@{}l@{}} conc-\\hashmap-\\put \end{tabular} &
\begin{tabular}[c]{@{}l@{}}Conc-\\urrentHash-\\Map.put\end{tabular}                                                                                         & 
\begin{tabular}[c]{@{}l@{}}
The rule detects code that calls `ConcurrentHashMap.containsKey()`, followed by a \\ call to `put()` if `containsKey` returned `False`. 
Since these operations are not atomic, \\ the atomicity violation can lead to a data loss.
\end{tabular}                                                                                                            \\\hline

10                                  & 
\begin{tabular}[c]{@{}l@{}} countdown-\\latch-await\end{tabular} &
\begin{tabular}[c]{@{}l@{}}Count-\\DownLatch.\\await\end{tabular}                                                                                          & 
\begin{tabular}[c]{@{}l@{}}
The `await` method returns `False` when the specified waiting time elapses while  \\ the thread is waiting for the latch to count down 
to zero. One must check the \\ value returned by `await` or check if `CountDownLatch.getCount() > 0` to handle  \\ the case
when the `await` call had timed out.
\end{tabular}                                                                                                     \\\hline

11                                  & 
\begin{tabular}[c]{@{}l@{}} create-list-\\from-map \end{tabular} &
\begin{tabular}[c]{@{}l@{}}  Map.\\values \end{tabular}        &
\begin{tabular}[c]{@{}l@{}}
The rule detects code that converts the `Collection` returned by `Map.values()` into  \\ a Java `Stream` and then collect it into a `List`. 
This can be simplified by calling  \\ the `ArrayList` constructor on the `Collection` returned by `values()`.\end{tabular}                                                                                                      \\\hline

12                                  & 
\begin{tabular}[c]{@{}l@{}} deprecated-\\base64       \end{tabular} &
\begin{tabular}[c]{@{}l@{}}Spring \\ Base64.\\encode\end{tabular}                                                                                           & 
\begin{tabular}[c]{@{}l@{}}
The `Base64` class in the Spring framework's crypto library is deprecated. Instead \\ of calling `Base64.encode()`, one must use 
`encodeToString` declared in the \\ `java.util.Base64.Encoder` class.\\ 
\textit{URL: \href{https://docs.spring.io/spring-security/site/docs/current/api/org/springframework/security/crypto/codec/Base64.html}{
\begin{tabular}[c]{@{}l@{}}
https://docs.spring.io/spring-security/site/docs/current/api/org/\\springframework/security/crypto/codec/Base64.html\end{tabular}
}}\end{tabular}                                                                                                   \\\hline

13                                  & 
\begin{tabular}[c]{@{}l@{}} deprecated-\\injectview  \end{tabular} &
\begin{tabular}[c]{@{}l@{}}  ButterKnife.\\inject   \end{tabular}        &                                                                                                                              
\begin{tabular}[c]{@{}l@{}}
ButterKnife InjectView was deprecated in version 7.  One must replace \\ `ButterKnife.inject(...)` with `ButterKnife.bind()` or use 
 Android's View Binding.\\
\textit{URL: \href{https://github.com/JakeWharton/butterknife/blob/master/CHANGELOG.md\#version-700-2015-06-27}{
\begin{tabular}[c]{@{}l@{}}
https://github.com/JakeWharton/butterknife/blob/master/CHANGELOG.md\\\#version-700-2015-06-27
\end{tabular} 
}}\end{tabular}                                                                                                                             \\\hline

14                                  &
 \begin{tabular}[c]{@{}l@{}}deprecated-\\mapping-\\ exception\end{tabular} 
& \begin{tabular}[c]{@{}l@{}}Jackson \\ mapping-\\Exception\end{tabular}                                                                                       & 
\begin{tabular}[c]{@{}l@{}}
Method `mappingException()` was deprecated in version 2.8 of the jackson-\\ databind library. Instead use `handleUnexpectedToken()`.\\
\textit{URL: \href{https://fasterxml.github.io/jackson-databind/javadoc/2.8/com/fasterxml/jackson/databind/DeserializationContext.html\#mappingException(java.lang.Class)}{
\begin{tabular}[c]{@{}l@{}}https://fasterxml.github.io/jackson-databind/javadoc/2.8/com/fasterxml/jackson/\\databind/DeserializationContext.html\#mappingException(java.lang.Class)\end{tabular}
}}\end{tabular}                         \\\hline

15                                  & 
\begin{tabular}[c]{@{}l@{}} deserialize-\\json-\\array   \end{tabular} &
 \begin{tabular}[c]{@{}l@{}}Gson.\\fromJson\end{tabular} &
\begin{tabular}[c]{@{}l@{}}
The rule detects code that deserializes a list of JSON items by iterating in a loop.  \\ Instead, one can directly deserialize into a list by 
specifying the correct parameterized\\  type using the `TypeToken` class.\\
\textit{URL: \href{https://github.com/google/gson/blob/master/UserGuide.md\#TOC-Serializing-and-Deserializing-Generic-Types}{
\begin{tabular}[c]{@{}l@{}}https://github.com/google/gson/blob/master/UserGuide.md\#TOC-Serializing-\\and-Deserializing-Generic-Types\end{tabular}
}}\end{tabular} \\\hline

16                                  & 

\begin{tabular}[c]{@{}l@{}} exception-\\invoke\end{tabular} &
\begin{tabular}[c]{@{}l@{}} Method.\\invoke  \end{tabular} &
 \begin{tabular}[c]{@{}l@{}}
Whenever `Method.invoke` is called, one must explicitly handle the `Invocation-\\TargetException`.  Since the actual underlying 
exception is the cause of `Invocation-\\TargetException`, it is desirable to call `Throwable.getCause()` or `getTargetException()` \\ 
in the   catch handler to access more information about the underlying exception.\end{tabular}               \\\hline

17                                  & 
\begin{tabular}[c]{@{}l@{}}executor-\\graceful-\\ shutdown\end{tabular}   & 
\begin{tabular}[c]{@{}l@{}}Executor-\\Service. \\shutdown-\\Now\end{tabular}                                                                                   & 
\begin{tabular}[c]{@{}l@{}}
One must shutdown an ExecutorService gracefully by first calling `shutdown` to  \\ reject any incoming tasks, waiting a while 
for the existing tasks to terminate by  \\ calling `awaitTermination`, and then calling `shutdownNow` to cancel \\  lingering tasks.  This is not required when the code calls `ExecutorService.\\invokeAll` that waits till all the tasks complete. \\
\textit{URL: \href{https://stackoverflow.com/questions/51819342}{https://stackoverflow.com/questions/51819342}}\end{tabular}                                                  \\\hline

18                                   & 
\begin{tabular}[c]{@{}l@{}}layout-\\inflater     \end{tabular} &
\begin{tabular}[c]{@{}l@{}}Android \\ TextView \\ constructor\end{tabular}                                                                                   & 
\begin{tabular}[c]{@{}l@{}}
One must inflate views using the `LayoutInflater` instead of creating `TextView`'s \\ programmatically in code.  Especially, if the 
layout is complex, it is much easier to define \\  it in XML and inflate it, rather than creating it all in code.\\
\textit{URL: \href{https://developer.android.com/reference/android/view/LayoutInflater}{https://developer.android.com/reference/android/view/LayoutInflater}}\end{tabular}                       \\\hline

19                                   & 
\begin{tabular}[c]{@{}l@{}} read-\\parcelable    \end{tabular} &
\begin{tabular}[c]{@{}l@{}}Parcel.\\readValue       \end{tabular}                                                                                   &                                                           
\begin{tabular}[c]{@{}l@{}}If one knows the specific type of the read object, one must use `readParcelable` \\ instead of calling `readValue` followed by an explicit  type-cast. Using `readParcelable` \\ will not require the type-cast operation.\end{tabular}                                                                                                                                                                       \\\hline

20                                  & 
\begin{tabular}[c]{@{}l@{}} replace-\\long-\\constructor \end{tabular} &
\begin{tabular}[c]{@{}l@{}}Long \\constructor    \end{tabular}                                                                                   &
\begin{tabular}[c]{@{}l@{}}Instead of constructing a new `Long` object, one must use `Long.valueOf()` as this  method \\ yields better space and time performance by caching frequently requested  values. \\
\textit{URL: \href{https://docs.oracle.com/javase/8/docs/api/java/lang/Long.html\#valueOf-long-}{https://docs.oracle.com/javase/8/docs/api/java/lang/Long.html\#valueOf-long-}}\end{tabular}                                                                                          \\\hline

21                                 & 
\begin{tabular}[c]{@{}l@{}} start-\\activity \end{tabular} &
\begin{tabular}[c]{@{}l@{}}Android\\ start-\\Activity\end{tabular}                                                                                           & 
\begin{tabular}[c]{@{}l@{}}
When launching an Android activity with `startActivity(Intent, ...)`, one must \\ check `Intent.resolveActivity(..)` for null. This checks 
if there exists an app on the  \\ device that can receive the implicit intent and launch the activity.  Otherwise, the  \\ app will crash  when 
`startActivity` is called.  This is not required when the activity  \\ is part of the same app,  or when `startActivity` is called within a try-catch block.\\
\textit{URL: \href{https://developer.android.com/guide/components/intents-common}{https://developer.android.com/guide/components/intents-common}}\end{tabular}                   \\\hline

22                                 & 
\begin{tabular}[c]{@{}l@{}} upgrade-\\enumerator    \end{tabular} &
\begin{tabular}[c]{@{}l@{}}Apache \\Enumerator \\ constructor\end{tabular}                                                                                  & 
\begin{tabular}[c]{@{}l@{}}
The rule detects code that constructs an object of the deprecated `Enumerator` class.  \\ One must use 
`Collections.enumeration` instead.\\
\textit{URL: \href{https://tomcat.apache.org/tomcat-7.0-doc/api/org/apache/catalina/util/Enumerator.html}{
\begin{tabular}[c]{@{}l@{}}
https://tomcat.apache.org/tomcat-7.0-doc/api/org/apache/catalina/util/Enumerator.html\end{tabular}
}}\end{tabular}                                                                                                                                                                        \\\hline

23                                  & 
\begin{tabular}[c]{@{}l@{}} upgrade-\\http-\\client \end{tabular} &
 \begin{tabular}[c]{@{}l@{}}HttpClient.\\ execute-\\Method\end{tabular}                                                                                       & 
\begin{tabular}[c]{@{}l@{}}
The rule detects code that calls `HttpClient.executeMethod`. This method has been  \\ replaced in the HttpClient library version 4 with 
method `execute`.  One  \\ must upgrade to the latest  version of the HttpClient library.\\
\textit{URL: \href{https://stackoverflow.com/questions/40795037}{https://stackoverflow.com/questions/40795037}}\end{tabular}                                                                                                                 \\\hline

24                                  & 
\begin{tabular}[c]{@{}l@{}} use-\\collectors-\\joining \end{tabular} &
 \begin{tabular}[c]{@{}l@{}} Stream.\\collect \end{tabular}                                                                                       &                                                
\begin{tabular}[c]{@{}l@{}}
The rule detects code that collects all the items into a `Collection` by calling \\ `Stream.collect()` and then joins them into a delimited 
String by `join()`. Instead,  \\ one can use `Collectors.joining(CharSequence)` to directly create a delimited String.\end{tabular}                                                                                           \\\hline

25                                  & 
\begin{tabular}[c]{@{}l@{}} use-file-\\read-\\utility\end{tabular} &
\begin{tabular}[c]{@{}l@{}}HttpURL-\\Connection\\ getInput-\\Stream\end{tabular}                                                                                & \begin{tabular}[c]{@{}l@{}}
The rule detects code that creates a Reader from a Connection's input stream and \\ reads its content by iterating in a loop. Instead, 
one can directly use a utility function,\\   such as `IOUtils.toString`, to read the input stream. This makes the  code 
more \\ readable and also prevents any chances of resource leaks.\end{tabular}                                                                                                                       \\\hline

26       &                           
\begin{tabular}[c]{@{}l@{}} use-fs-\\is-dir \end{tabular} &                                                                                                                                                                               
\begin{tabular}[c]{@{}l@{}}FileSystem.\\ isDirectory\end{tabular}                                                                                         &
\begin{tabular}[c]{@{}l@{}}
`FileSystem.isDirectory()` has been deprecated. One must use   `FileSystem.\\getFileStatus().isDir()` instead. \\
\textit{URL: \href{https://hadoop.apache.org/docs/current/api/org/apache/hadoop/fs/FileSystem.html\#isDirectory-org.apache.hadoop.fs.Path-}{\begin{tabular}[c]{@{}l@{}}https://hadoop.apache.org/docs/current/api/org/apache/hadoop/fs/\\FileSystem.html\#isDirectory-org.apache.hadoop.fs.Path-\end{tabular}}}\end{tabular}                                                                                                               \\\hline

27         &                         
\begin{tabular}[c]{@{}l@{}} use-\\guava-\\hashmap\end{tabular} &                  
\begin{tabular}[c]{@{}l@{}} HashMap \\constructor \end{tabular}                                                                                         &                                                                                                                                                                      
\begin{tabular}[c]{@{}l@{}}
The rule detects code that constructs a HashMap of a given size and then \\ immediately adds all the keys into the map by iterating 
in a loop. The HashMap \\ constructor uses a default load factor of 0.75, which means that the hash table \\ will be rehashed after 75\% of all
keys have been added to the table.  One must use \\ Guava's `newHashMapWithExpectedSize()` as that will not result in a rehash. \\
\textit{URL: \href{https://stackoverflow.com/questions/30220820}{https://stackoverflow.com/questions/30220820}}\end{tabular}                                            \\\hline

28          &                         
\begin{tabular}[c]{@{}l@{}} use-\\parcelable\end{tabular} &                                                                                                                       
\begin{tabular}[c]{@{}l@{}}Android\\ getSerial-\\izableExtra\end{tabular}                                                                                    & 
\begin{tabular}[c]{@{}l@{}}One must use `Parcelable` instead of `Serializable` to pass data between  different \\components in Android, as the former is more performant. \\
\textit{URL: \href{https://skoric.svbtle.com/serializable-vs-parcelable}{https://skoric.svbtle.com/serializable-vs-parcelable}}\end{tabular}                                                                        \\\hline

29                                  & 
\begin{tabular}[c]{@{}l@{}} use-\\remove-if\end{tabular} &           
\begin{tabular}[c]{@{}l@{}} Iterator.\\remove\end{tabular}                                                                                    &                                                   
\begin{tabular}[c]{@{}l@{}}
The rule detects code that conditionally removes values from a hashmap by iterating \\  over all the map entries in a loop. One can instead use  ‘Collection.removeIf‘. \\ This is also more efficient than removing values  by iterating over all entries.
\end{tabular} \\\hline

30        &                           
\begin{tabular}[c]{@{}l@{}} view-\\binding \end{tabular} &                                                             
 \begin{tabular}[c]{@{}l@{}}Android \\ findViewById\end{tabular}                                                                                           & 
\begin{tabular}[c]{@{}l@{}}One must enable view binding in their module instead of calling `findViewById()`. \\ View binding provides null-safety and type-safety at compile time. \\
\textit{URL: \href{https://developer.android.com/topic/libraries/view-binding\#java}{https://developer.android.com/topic/libraries/view-binding\#java}}\end{tabular}                                                                                                                             \\\hline

31                                 & 
\begin{tabular}[c]{@{}l@{}} wrapping-\\exception \end{tabular} & 
\begin{tabular}[c]{@{}l@{}} Invocation-\\TargetException.\\getCause \end{tabular} & 
\begin{tabular}[c]{@{}l@{}}
The rule detects code that wraps `InvocationTargetException.getCause()` into an \\ unchecked exception.  One must check if the `Throwable` returned by `getCause()` \\ can be itself type-cast into an unchecked exception.  A new unchecked  exception \\ must be  constructed only if the type-casting operation is not successful.
\end{tabular} \\\hline
\caption{\label{fig:bp_description} The synthesized rules along with pointers to supporting blog-posts and code documentation.} 
\end{tabularx}
\end{center}

\begin{table*}[]
\resizebox{0.55\textwidth}{!}{

\begin{tabular}{l c c c c}
Rule name                                                                                & \begin{tabular}[c]{@{}l@{}}\# code \\ changes\end{tabular} & \begin{tabular}[c]{@{}l@{}}\# new \\ egs. \end{tabular} & 
\begin{tabular}[c]{@{}l@{}} PreC \\ size\end{tabular} & \begin{tabular}[c]{@{}l@{}} PostC \\ size\end{tabular} \\ \hline

check-file-rename & 6 & - & 4 & -\\
check-inputstream-skip & 5 & - & 3 & -\\
check-mkdirs & 10 & - & 1 & -\\
check-resultset-next & 4 & - & 8 & -\\
conc-hashmap-put & 3 & - & 9 & -\\
create-list-from-map & 3 & - & 8 & -\\
deprecated-base64 & 4 & - & 4 & -\\
deprecated-injectview & 16 & - & 6 & -\\
\begin{tabular}[c]{@{}l@{}}deprecated-mapping-\\ exception\end{tabular} & 5 & - & 8 & -\\
deserialize-json-array & 3 & - & 10 & -\\
layout-inflater & 3 & - & 5 & -\\
read-parcelable & 7 & - & 6 & -\\
replace-long-constructor & 4 & - & 4 & -\\
upgrade-http-client & 3 & - & 7 & -\\
upgrade-enumerator & 6 & - & 6 & -\\
use-collectors-joining & 3 & - & 12 & -\\
use-file-read-utility & 3 & - & 15 & -\\
use-fs-is-dir & 4 & - & 4 & -\\
use-guava-hashmap & 3 & - & 20 & -\\ 
use-parcelable & 3 & - & 4 & - \\
use-remove-if & 3 & - & 11 & - \\ 
view-binding & 3 & - & 5 & -\\

wrapping-exception & 7 & - & 6 & 8 \\ \hline

\multicolumn{5}{c}{Rules Iteratively Refined} \\ \hline
check-actionbar & 3 & 3 & 10 & (14, 14)\\
\begin{tabular}[c]{@{}l@{}}{check-await-}\\ termination\end{tabular} & 5 & 2 & 8 & 12 \\
check-createnewfile & 3 & 3 & 4 & 6\\
check-movetofirst & 4 & 9 & 2 & (6, 8)\\
countdownlatch-await & 11 & 3 & 4 & 9\\
exception-invoke & 4 & 3 & 3 & 6\\
\begin{tabular}[c]{@{}l@{}} {executor-graceful}-\\ {shutdown} \end{tabular} & 4 & 4 & 2 & (5, 10)\\
{start-activity} & 5 & 5 & 3 & (5, 6, 11)\\
\hline
\end{tabular}
}	
        \caption{Information about the synthesized rules. \# new egs. are additional examples used for rule refinement,  PreC size and  PostC size are the number of nodes in the rule precondition and (possibly disjunctive) postcondition formulas. $(i_1, i_2,\cdots)$ indicates a postcondition with disjuncts of sizes $i_1$, $i_2$, $\cdots$ respectively.}
        \label{fig:rule_synthesis_stats}              
\end{table*}

\begin{table*}[]
\resizebox{0.6\textwidth}{!}{

\begin{tabular}{l ccc ccc}
\multirow{2}{*}{Rule name} & \# code &  \multicolumn{2}{c}{Before rule} & \# new & \multicolumn{2}{c}{After rule}  \\
& changes & \multicolumn{2}{c}{refinement}  & egs. & \multicolumn{2}{c}{refinement} \\ \cline{3-4} \cline{6-7}

& &  PostC & Prec. & &  PostC & Prec. \\
& & size &  & & size & \\ \hline

\begin{tabular}[c]{@{}l@{}} check-\\actionbar  \end{tabular} & 	3	& 14 & 	32\%	 & 3	& (14, 14)	& 100\% \\ \hline 
\begin{tabular}[c]{@{}l@{}}check-await-\\ termination\end{tabular} & 5& -& 79\% & 2 & 12 & 100\%\\\hline
\begin{tabular}[c]{@{}l@{}} check-create\\newfile	 \end{tabular} & 3 & -	& 42\%	& 3	& 6	& 100\% \\\hline
\begin{tabular}[c]{@{}l@{}} check-\\movetofirst	  \end{tabular} & 4 & -	& 40\% &	9	& (6, 8) &	100\%  \\\hline
\begin{tabular}[c]{@{}l@{}} countdown\\latch-await	 \end{tabular} & 11 & -	& 80\%	& 3	& 9	& 100\% \\\hline
\begin{tabular}[c]{@{}l@{}} exception-\\invoke	 \end{tabular} & 4	& 9 &	76\% &	3	& 6 &	100\% \\\hline
\begin{tabular}[c]{@{}l@{}} executor-graceful-\\shutdown \end{tabular} & 4& 9 & 95\%& 4 & (5, 10) & 100\% \\ \hline
start-activity & 5& 9& 21\%& 5 & (5, 6, 11) & 80\% \\ \hline
Macro Avg. &  & 	& 58\% &		&  &	97\%
\end{tabular}
}
\caption{\label{fig:iterative_refinement} Results for iterative rule refinement (PostC size is the number of nodes in each disjunct of the synthesized postcondition, Prec. is the precision based on user-study).}
\vspace{-0.6cm}
\end{table*}

\subsection{Effectiveness of Iterative Rule Refinement}
\label{sec:rq2}

We iteratively refine $8$ rules by providing additional non-buggy code examples. These examples 
correspond to variations in correct code that are not present in the code changes. 
As can be seen from Table~\ref{fig:iterative_refinement}, few additional examples are sufficient to refine the rule. All the 8 rules show precision improvements and 
the (macro) average precision increases from $58\%$ to $97\%$ based on refinement.
We use labels from offline evaluation for estimating the precision
\footnote{We group all detections by the
rule version, and estimate the overall precision based on 10 
labels for each group.}.
In 4 of the 8 rules, the refinement occurs by synthesizing disjunctive postconditions.

\begin{table*}[]
\resizebox{0.45\textwidth}{!}{

        \begin{tabular}{l l c}
        
        Rule name & 
        \begin{tabular}[c]{@{}l@{}}ProSynth\end{tabular}            & 
        \begin{tabular}[c]{@{}l@{}}AST anti-\\unification\end{tabular}        \\\hline

deprecated-injectview \\
\begin{tabular}[c]{@{}l@{}}deprecated-mapping-\\ exception\end{tabular}            \\ 
layout-inflater	\\
replace-long-constructor \\        
use-parcelable \\
view-binding \\
upgrade-enumerator\\

check-actionbar  & \xmark, TO \\
check-createnewfile  & \xmark, TO \\
check-file-rename  & \xmark, TO \\
check-inputstream-skip & \xmark, TO \\
check-movetofirst & \xmark, TO \\
conc-hashmap-put & \xmark, MC \\
create-list-from-map & \xmark, MC \\
deprecated-base64 & \xmark, MC \\
read-parcelable & \xmark, MC \\
upgrade-http-client & \xmark, MC \\
use-fs-is-dir & \xmark, MC \\ 
wrapping-exception & \xmark, TO  \\

\begin{tabular}[c]{@{}l@{}}check-await-\\ termination\end{tabular} & \xmark, TO & \xmark \\
check-mkdirs & \xmark, TO & \xmark \\
check-resultset-next & \xmark, TO & \xmark \\
countdownlatch-await  & \xmark, UnSAT & \xmark \\
deserialize-json-array & \xmark, MC & \xmark \\
exception-invoke & \xmark, TO & \xmark \\
\begin{tabular}[c]{@{}l@{}}executor-graceful-\\ shutdown\end{tabular} & \xmark, TO & \xmark \\
start-activity & \xmark, TO & \xmark\\
use-collectors-joining  & \xmark, MC & \xmark \\
use-file-read-utility & \xmark, MC & \xmark\\
use-guava-hashmap & \xmark, MC & \xmark \\       
use-remove-if & \xmark, MC & \xmark \\       \hline
        \end{tabular}
}
\caption{Comparison with ProSynth and an AST anti-unification based approach. In the above: TO means timeout at 15m, MC means missing code context, UnSAT means unsatisfiable problem.}
\label{fig:baselines}              
\end{table*}

\subsection{Comparison with baselines}
\label{sec:rq3}

We compare \RhoSynth\ with ProSynth~~\cite{provenance} and AST anti-unification based approaches, on synthesizing rules without the rule refinement step.  Getafix~\cite{getafix} and Revisar~\cite{revisar} are representatives for synthesizing AST transformations via anti-unification of tree patterns. A direct comparison against Getafix and Revisar is unfortunately not feasible as the former is not publicly available and the latter does not support Java. Hence, we compare against our own implementation of an anti-unification algorithm over ASTs. The results are described in Table~\ref{fig:baselines}. We  provide artifacts from these experiment as supplementary material.

\medskip\noindent \textbf{ProSynth:} 
ProSynth is a general algorithm to synthesize Datalog programs. 
It is not particularly tailored to the rule synthesis problem, which is the focus of our work, for two main reasons. 
First, a datalog program can only express rules that are purely existentially quantified formulas without a postcondition~\cite{datalog-fol}. 
ProSynth is thus not able to synthesize rules that have a postcondition. It either times out or returns "Problem unsatisfiable" when synthesizing such rules.  

Second, in cases when ProSynth is able to synthesize a rule, the synthesized rule often misses the required code context, i.e., the API of interest or constructs of interest.  
Synthesizing a rule from few examples is in most cases an underspecified problem.  While \RhoSynth\ biases the synthesis of rule preconditions towards larger code context,  ProSynth uses a SAT solver based enumeration to synthesize a rule precondition without such an inductive bias.  
As a result, rules synthesized by ProSynth can miss the required code context as long as they can differentiate between the few positive and negative code examples
\footnote{
As an example, the code context for \texttt{\footnotesize conc-hashmap-put} rule must check the result of \texttt{\footnotesize ConcurrentHashMap.containsKey} followed by a call to \texttt{\footnotesize put}, on the same hash map, if \texttt{\footnotesize containsKey} returned \texttt{\footnotesize False}.  The rule synthesized by ProSynth just checks for an existence of a \texttt{\footnotesize containsKey} call and misses the remaining code context.}.

ProSynth times out on 12 rules and returns "Problem unsatisfiable" for 1 rule.  On a manual examination of the 18 rules it synthesized,  we found that $11$ rules do not contain the required code context.  These rules would lead to false positives.  Hence, we conclude that ProSynth is able to precisely synthesize $7$ out of $31$ rules.

\medskip\noindent\textbf{Anti-unification over ASTs:} 
Given two ASTs, we look at all combination of subtrees rooted at different nodes in the two ASTs. We pick the AST pattern, obtained on anti-unification of these subtrees, that has the  largest size and contains the API of interest (column $3$ in Table~\ref{fig:bp_description}). 
We manually examine the AST patterns obtained on anti-unifying \texttt{\small code-before} and \texttt{\small code-after} examples. We observe that due to variations in the syntactic structure of code changes, anti-unification misses the required code context in $12 /31$ rules. These AST patterns would lead to false positives. 
 This approach is thus able to precisely synthesize at most $19$ rules. This experiment shows the limitation of expressing rules over an AST representation.

\begin{table*}[]
\resizebox{0.95\textwidth}{!}{
\begin{tabular}{lc c c ccc}
\begin{tabular}[c]{@{}l@{}} Iteratively\\refined rule\end{tabular} & 
\begin{tabular}[c]{@{}l@{}} PostC\\disjunct\end{tabular} & 
\begin{tabular}[c]{@{}l@{}} \# new \\ unpaired \\ egs.\end{tabular} &
\begin{tabular}[c]{@{}l@{}} \# action \\ nodes in \\ rule PostC \end{tabular} &
\begin{tabular}[c]{@{}l@{}} \# Node mappings \\~~~using ILP graph- \\ alignment \end{tabular} & 
\begin{tabular}[c]{@{}l@{}} \#Node mappings\\ using GumTree \end{tabular} & 
\begin{tabular}[c]{@{}l@{}}  PostC synthesis \\~~~succeeds\\ with GumTree? \end{tabular} \\ \hline

\multirow{2}{*}{check-actionbar}  & $\textit{post}_1$ & - & 6 &- & -& \\
						& $\textit{post}_2$ & 3 & 6 & 18/18 = 100\% & 18/18 = 100\% & \cmark \\ \hline

check-await-termination &  $\textit{post}_1$ & 2 & 6 & 6/6 = 100\% & 4/6 = 66\% & \cmark \\ \hline

check-createnewfile  & $\textit{post}_1$ & 3 & 4 & 12/12 = 100\% & 7/12 = 58\% & \xmark \\ \hline

\multirow{2}{*}{check-movetofirst}  & $\textit{post}_1$  & 6  & 3 & 45/45 = 100\% & 41/45 = 91\% & \xmark \\
							& $\textit{post}_2$    & 3 & 4 & 12/12 = 100\% & 10/12= 83\% & \cmark \\ \hline

countdownlatch-await  & $\textit{post}_1$ & 3 & 5 & 15/15 = 100\% & 7/15 = 47\% & \xmark \\ \hline

exception-invoke  & $\textit{post}_1$ & 3 & 3 & 9/9 = 100\% & 3/9 = 33\% & \xmark \\ \hline

\multirow{3}{*}{executor-graceful-shutdown} &  $\textit{post}_1$ & 1 & 2 & - &-  & \\ 
								     &  $\textit{post}_2$ & 3 & 6 & 18/18 = 100\% & 9/18 = 50\% &\xmark \\\hline

\multirow{3}{*}{start-activity} &  $\textit{post}_1$ & 3 & 2 & 6/6 = 100\% & 6/6 = 100\% & \cmark \\
					     &  $\textit{post}_2$ & 1 & 3 & - & - & \\
					     &  $\textit{post}_3$ & - & 5 & - &-  & \\\hline

\end{tabular}
}
\caption{\label{fig:ilp} Comparison of ILP based graph alignment with GumTree on aligning unpaired code examples for rule synthesis.
Node mappings are only reported for those partitions of conforming examples that contain at least two new unpaired examples.  We compare alignment over just action nodes since they have a 1-1 map between ASTs and PDGs. }
\end{table*}

\subsection{ILP-based Graph Alignment vs GumTree}
\label{sec:rq4}

We now compare ILP-based alignment and state-of-the-art code differencing algorithms on \textit{aligning unpaired code examples for rule synthesis}.
GumTree~\cite{gumtree} is a popular choice for performing AST differencing. It is effective when aligning code changes in which the two snippets overlap a lot. If the examples are ``unpaired'' (or have small overlap), GumTree fails in some cases to precisely align them. 
We consider $9$ scenarios covering all $8$ iteratively refined rules where a conjunctive postcondition formula is synthesized from a partition of unpaired conforming code examples. 
In Table~\ref{fig:ilp}, we tabulate the number of nodes mappings established by the two  alignment algorithms for every pair of code example in the partition.
The number of nodes mapped by GumTree range from $33\%$ to $91\%$ in the failed scenarios.  ILP maps all the nodes.

When we use node mappings established by the GumTree algorithm to \texttt{\small merge} these  examples, we succeed in synthesizing the desired postcondition  in only $4 / 9$ scenarios.
In the remaining scenarios, GumTree fails to map some nodes that form the required context for the rule postcondition, and are hence omitted on subsequent merge operations. These rules would lead to false positives. 
The ILP-formulation leads to successful synthesis of the specific postcondition in all $9$ scenarios.

\section{Related Work} \label{relwork:sec}


\noindent\textbf{Synthesis algorithms for program repair and bug detection:} 
Our work is most closely related to research on synthesis algorithms for automated program repair from code changes~\cite{getafix,revisar,refazer,bluepencil,phoenix,lase}. We can categorize these works based on their techniques. The first category consists of Phoenix~\cite{phoenix} and instantations of the PROSE framework~\cite{prose} in Refazer~\cite{refazer} and BluePencil~\cite{bluepencil}. These algorithms search for a program transformation, which explain the set of provided concrete edits, within a DSL. The second category includes Revisar~\cite{revisar}, Getafix~\cite{getafix} and LASE~\cite{lase} that are based on generalization algorithms over ASTs, using anti-unification or maximum common embedded subtree extraction.
Both the above categories of work either rely on an accompanying static analyzer for bug localization or synthesize program transformation patterns that are specific to a given package or domain. 
Bug localization is harder than repair at a given location~\cite{self_supervised_bug_fix}.  
Our approach does not rely on existing static analyzers for bug localization and is able to precisely synthesize new rules. These rules are also applicable to code variations that are present in different packages. 
Additionally, we observe that correct code may have variations that is not present in code changes. We present an approach to refine rules using additional examples containing such code variations. In comparison, most of the above mentioned prior works rely on paired code changes and none of them support iterative refinement of rules.

DiffCode~\cite{deepcode-pldi18} is an approach to infer rules from code changes geared towards Crypto APIs. Their main focus is on clustering algorithms for code changes.  
They do not automate the task of synthesizing a  rule from a cluster of code changes.
Program synthesis has also been recently applied to other code related applications such as API migration~\cite{api_migration,apifix}, synthesis of merge conflict resolutions~\cite{conflict_resolution}, interactive code search~\cite{sporq}.

\noindent\textbf{Datalog synthesis: }
There is a  rich body of recent work on datalog synthesis~\cite{provenance,alps,difflog,zaatar,gensynth,egs} and  its application to code related tasks such as interactive code search~\cite{sporq}. Datalog programs can express existentially quantified preconditions~\cite{datalog-fol} but cannot express rules with a non-vacuous postcondition that introduce quantifier alternation. 
On the other hand, these algorithms can synthesize recursive predicates that we exclude in our approach. In Section~\ref{expt:sec}, we provide an empirical comparison against ProSynth~\cite{provenance} from this category.

\noindent\textbf{Statistical approaches for program synthesis: }
\cite{hoppity,tufano} are neural approaches for automatic bug-fix generation. 
These models are trained on a general bug-fix dataset, not necessarily fixes that correspond to  rules, and have a comparatively lower accuracy. TFix~\cite{deepcode-bugfix} improves upon them by fine-tuning the neural models on 
fixes that correspond to a known set of bug categories or rules.  
In doing so, it gives up the ability to generate fixes for new rules.
~\cite{rishabh_singh_ssc,prophet} propose a hybrid approach where algorithmic techniques are used to generate candidate fixes and statistical models are used to rank them.
~\cite{ggnn,deep-bugs,vasic2018neural,self_supervised_bug_fix} present neural models for detecting bugs caused by issues in variable naming and variable misuse. These approaches cannot be easily tailored to generate new detectors from a given set of labeled code examples. 

In the domain of synthesizing String transformations, pioneered by FlashFill~\cite{flashfill}, RobustFill~\cite{robustfill} presents a neural approach for program synthesis as well as program induction. 
Hybrid approaches such as~\cite{pl-meets-ml,neural-guided-search,semantic_pbe} complement machine learning based induction with algorithmic techniques and present an interesting direction for exploration in the context of rule synthesis.

Statistical approaches for building bug-detectors include data-mining approaches such as APISan~\cite{apisan}, PR-Miner~\cite{prminer} and NAR-miner~\cite{narminer}. 
These approaches mine popular programming patterns and flag deviants as bugs. Since these mined patterns are based on frequency, these approaches are not able to distinguish between incorrect code and infrequent code. Recent work on mining code patterns has achieved higher precision in finding bugs when the mined patterns are applied to the same code-base they are extracted from~\cite{code-inconsistency}. 
Arbitrar~\cite{arbitrar} improves upon APISan~\cite{apisan} by incorporating user-feedback through active learning algorithms.  Both Arbitrar and APISan employ symbolic execution  to extract semantic features of a program,  as opposed to a lighter-weight static analysis in \RhoSynth.  Rules based on symbolic execution might be too expensive for a real-time code reviewing application.

\noindent\textbf{Interactive program synthesis}: 
Researchers have recently explored the question-selection problem~\cite{question_selection} and other user-interaction models~\cite{interactive_prog_synthesis,gauss,loopy} in the context of interactive program synthesis.
Rule refinement is also an instance of interactive program synthesis and applying these approaches to the domain of rule synthesis is an interesting research direction.

\noindent\textbf{DSLs for expressing rules in static checkers: }
Most static checkers, such as SonarQube\footnote{\url{https://docs.sonarqube.org/latest/extend/adding-coding-rules/}}, PMD\footnote{\url{https://pmd.github.io/latest/pmd_userdocs_extending_writing_pmd_rules.html}}, Semmle and Semgrep, allow users to write their own custom rules. These rules are often written in a DSL such as ProgQuery~\cite{progquery} or CodeQL.
We are not aware of any tool that synthesizes rules in these DSLs automatically from labeled code examples. In fact, a recent survey~\cite{deepcode-survey} identifies automating the rule creation process in static checkers as a largely unexplored area. 

\noindent\textbf{Synthesis frameworks: }
SyGus~\cite{sygus} and CEGIS~\cite{cegis} are two popular synthesis frameworks in domains with logical specifications.
Recently, Wang et al.~\cite{side-channel-synthesis} have proposed an approach that combines SyGus with decision tree learning for synthesizing specific static analyses that detect side-channel information leaks.

\section{Conclusions} \label{conc:sec}

In this paper, we present \RhoSynth, a new algorithm for synthesizing code-quality rules from labeled code examples. 
\RhoSynth\ performs rule synthesis on graph representations of code and is based on a novel ILP based graph alignment algorithm.
We validate our algorithm by synthesizing more than 30 rules that have been deployed in production. 
Our experimental results show that the rules synthesized by our approach have high precision (greater than 75\% in live code reviews) which make them suitable for real-world applications. In the case of low-precision rules, we show that rule refinement can leverage additional examples to incrementally improve the rules. 
Interestingly, these synthesized rules are capable of enforcing several documented code-quality recommendations. 
Through comparisons with recent baselines, we show that current state-of-the-art program synthesis approaches are unable to synthesize most rules.

\bibliography{references}

\clearpage

\appendix

\section{Input examples to \RhoSynth} \label{input:sec}
We obtain code changes from $27,752$ Java GitHub packages that were selected based on their license, i.e., Apache or MIT-license, and star rating\footnote{A list of these GitHub packages is included in the supplementary material.}. These code changes are extracted using CPatMiner~\cite{cpatminer}. The corpus consists of $1.84M$ code changes. 
We cluster PDG representation of these code changes in two phases -- (1) we
map each code change to a set of APIs that are added, deleted or replaced in the change; (2) for each API, we cluster the mapped code changes using a bottom-up hierarchical clustering algorithm~\cite{clustering_methods} (a code change is featurized using a fixed set of predicate features over its PDG representation and Jaccard distance is used to compute similarity between code changes).
The clustering algorithm is tuned to output clusters that have a high homogeneity score.  This ensures that all code-changes within the same cluster correspond to the same rule.

We tag each code change with the relative file path, the name and the line number of the method containing the change. We use this triple as  a criterion to remove code-change overlap within a cluster.
We filtered clusters containing less than 3 examples and having examples from less than 2 repositories or 2 commit ids. The latter ensures that the code changes are not repository-specific.
We then choose clusters containing popular APIs belonging to popular frameworks such as Java.util, Android, Apache, Guava, etc. The popularity is determined based on frequency of occurrence in the corpus. This results in 1397 clusters. We pick 300 clusters randomly from this set. The authors manually examined these clusters and select those clusters whose underlying APIs and code constructs find relevant results on a Google search, which is evidence for a general applicability of the code changes in the cluster. 
For e.g., code-changes for the “use-guava-hashmap” rule involve the \texttt{\small HashMap constructor} and \texttt{\small Maps.newHashMapWithExpectedSize} API that have an associated Stack-Overflow post. We spend on average ~5 minutes for manually examining each cluster. This examination resulted in 31 clusters which are then input to \RhoSynth.

The paper containing the details of clustering is under submission.

\section{Example of PDG}

\begin{figure}[h]
\centering
\resizebox{0.2\textwidth}{!}{
   \includegraphics[width=\columnwidth]{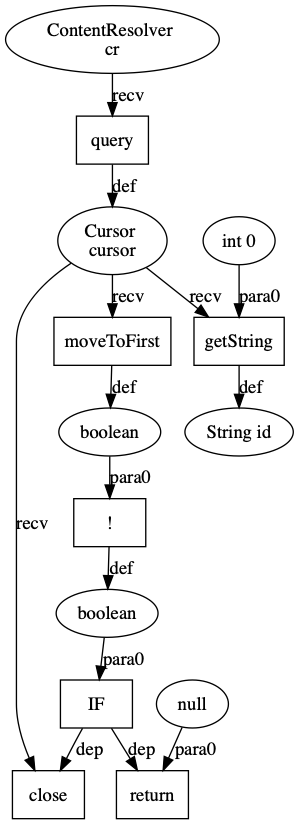}
}
   \caption{PDG for the snippet of code-after in Figure~\ref{fig:examples}(a).}
   \label{fig:pdg} 
\end{figure}

Refer to Figure~\ref{fig:pdg} for the PDG of the code-after snippet in Figure~\ref{fig:examples}(a). Oval nodes in the figure indicate data nodes and rectangular nodes indicate action nodes.

\label{appendix:pdg}

\section{Maximal Graph Alignment using ILP}
\label{app:ilp}
In this section, we describe the reduction from maximizing  graph alignment to an integer linear program (ILP). 
Let $\mathscr{A}_1 = (N_1, E_1, Lat_1, M_{f1}, M_{b1})$ and $\mathscr{A}_2 = (N_2, E_2, Lat_2$, $M_{f2}, M_{b2})$ be two UAPDGs over the same set of free variables $\textit{Var}_f$.
Graph alignment on $\mathscr{A}_1$
 and $\mathscr{A}_2$ outputs a node mapping $NM \subseteq N_1 \times N_2$ and an edge mapping $EM: E_1 \times E_2$ as follows.
The ILP optimization problem is defined over the following set of variables: 
\begin{enumerate} 
\item $z_{n_1n_2}$ are  0-1 variables that are set if $(n_1, n_2) \in NM$
\item $z_{e_1e_2}$ are 0-1 variables that are set if $(e_1, e_2) \in EM$
\item $z_{n_1}, z_{n_2}, z_{e_1}, z_{e_2}$ are 0-1 variables that are set if the corresponding nodes and edges are not mapped by $NM$ and $EM$ respectively
\end{enumerate}

The ILP objective function seeks to maximize the number of mapped nodes and edges:
$$\textit{Objective function} = \max\limits_{\stackrel{z_{n_1n_2}, z_{e_1e_2}, }{z_{n_1}, z_{n_2},z_{e_1},z_{e_2}}} \sum\limits_{n_1 \in N_1, n_2 \in N_2} z_{n_1n_2} + \sum\limits_{e_1 \in E_1, e_2 \in E_2} z_{e_1e_2}$$ 

The optimization is subject to the following constraints:

	\begin{enumerate}
				\item Each node $n_1 \in N_1$ is optionally mapped to a node in $N_2$: $z_{n_1} + \sum\limits_{n_2 \in N_2} z_{n_1n_2} = 1$ for all $n_1 \in N_1$.
			\item Each node $n_2 \in N_2$ is optionally mapped to a node in $N_1$: $z_{n_2} + \sum\limits_{n_1 \in N_1} z_{n_1n_2} = 1$ for all $n_2 \in N_2$.
			\item Each edge $e_1 \in E_1$ is optionally mapped to an edge in $E_2$: $z_{e_1} + \sum\limits_{e_2 \in E_2} z_{e_1e_2} = 1$ for all $e_1 \in E_1$.
			\item Each edge $e_2 \in E_2$ is optionally mapped to an edge in $E_1$: $z_{e_2} + \sum\limits_{e_1 \in E_1} z_{e_1e_2} = 1$ for all $e_2 \in E_2$.
		\item Nodes mapped to the same free variables are aligned: $z_{n_1n_2} = 1$ if $M_{f1}(n_1) = M_{f2}(n_2)$.
			\item Action nodes with different labels are not aligned: $z_{n_1n_2} = 0$ if $n_1$ and $n_2$ are action nodes with different labels.
		\item Action nodes with different change tags are not aligned while synthesizing preconditions: $z_{n_1n_2} = 0$ if change tag of $n_1$ and $n_2$ are different, where $n_1$, $n_2$ are both action nodes.
		\item Edges with different labels are not aligned: $z_{e_1e_2} = 0$  if $e_1$ and $e_2$ have different labels.
			\item Node and edge mappings satisfy the topological constraints. This means if edges $e_1$ and $e_2$ are mapped then the nodes at their sources and destinations must be mapped:
			$z_{e_1e_2} \leq z_{n_1n_2}$, where $n_1 = \textit{src}(e_1)$ and $n_2 = \textit{src}(e_2)$  for all $e_1 \in E_1$ and $e_2 \in E_2$. Similarly, $z_{e_1e_2} \leq z_{n_1n_2}$, where $n_1 = \textit{dest}(e_1)$ and $n_2 = \textit{dest}(e_2)$.
			\item If two data nodes align then they must have at least one aligned incoming or outgoing edge:\\ $z_{n_1n_2} \leq \sum\limits_{n_1=\textit{src}(e_1), n_2 = \textit{src}(e_2)} z_{e_1e_2} + \sum\limits_{n_1 = \textit{dest}(e_1, n_2 = \textit{dest}(e_2)} z_{e_1e_2}$ for data nodes $n_1$ and $n_2$.

	\end{enumerate}

\section{Proof of Theorem~\ref{thm:merge_soundness}}
\label{app:proof1}

We first introduce two new lemmas that help us in proving the main theorem.

\begin{lemma}
\label{lemma:projection}
Given a UAPDG $\mathscr{A} = (N, E, Lat, M_{f}, M_{b})$. Let $\mathscr{A}' = (N', E', Lat', M_{f}', M_{b}')$ be obtained by projecting $\mathscr{A}$ to a subset of nodes and edges, $N' \subseteq N$ and $E' \subseteq E$, in the following manner:
\begin{itemize}
\item[(a)] if $n \in M_f \Rightarrow n \in M_f'$.
\item[(b)] if $(n_1, n_2) \in E \textit{ and } n_1 \in M_f \textit{ and } n_2 \in M_f \Rightarrow (n_1,n_2) \in E'$.
\end{itemize}
Then $\mathscr{A} \Rightarrow \mathscr{A}'$.
\end{lemma}

\begin{proof}
Let $\mathscr{A}$ correspond to the formula $\exists \vec{y}. \varphi(\vec{x}, \vec{y})$ and $\mathscr{A'}$ correspond to the formula $\exists \vec{y}. \varphi'(\vec{x}, \vec{y})$. 
One can easily argue that $\varphi(\vec{x}, \vec{y}) \Rightarrow \varphi'(\vec{x}, \vec{y})$. Therefore, any model $A^{\vec{x}, \vec{y}}$ of $\varphi(\vec{x}, \vec{y})$ will also be a model of $\varphi'(\vec{x}, \vec{y})$. 
\end{proof}

\begin{lemma}
\label{lemma:restricted_soundness}
Let $\mathscr{A}_1 = (N_1, E_1, Lat_1, M_{f1}, M_{b1})$ and $\mathscr{A}_2 = (N_2,$ $E_2,$ $Lat_2$, $M_{f2}, M_{b2})$ be two UAPDGs defined over the same set of free variables $\vec{x} = \{x_1, \cdots, x_n\}$ and bound variables $\vec{y} = \{y_1, \cdots, y_k\}$. 
Let $NM \subseteq N_1 \times N_2$ and $EM \subseteq E_1 \times E_2$
be node and edge mappings obtained from aligning $\mathscr{A}_1$ and $\mathscr{A}_2$ such that
(a) all nodes in $N_1$ and $N_2$ and all edges in $E_1$ and $E_2$ are mapped through $NM$ and $EM$ respectively,
and 
(b)  $(n_1, n_2) \in NM \Rightarrow$ either $M_{f1}(n_1) = M_{f2}(n_2)$ or $M_{b1}(n_1) = M_{b2}(n_2)$.
Then $\mathscr{A}_1 \vee \mathscr{A}_2 \Rightarrow \texttt{merge}_{(NM, EM)}(\mathscr{A}_1, \mathscr{A}_2)$.
\end{lemma}

\begin{proof}
Let $\mathscr{A}(N, E, Lat, M_{f}, M_{b}) = \texttt{merge}_{(NM, EM)}(\mathscr{A}_1, \mathscr{A}_2)$.
Without loss of generality, let $M_b$ be such that $M_b(n_1, n_2) = M_{b1}(n_1) = M_{b2}(n_2)$. 
Let the UAPDGs $\mathscr{A}, \mathscr{A}_1$ and $\mathscr{A}_2$ correspond to formulas 
$\exists \vec{y}. \varphi(\vec{x}, \vec{y})$, $\exists \vec{y}. \varphi_1(\vec{x}, \vec{y})$ and $\exists \vec{y}. \varphi_2(\vec{x}, \vec{y})$ respectively.
Let $A_1^{\vec{x}, \vec{y}}$ be a graph structure with its nodes labeled with $\vec{x}$ and 
$\vec{y}$ such that $A_1^{\vec{x}, \vec{y}}$ is a model for $\mathscr{A}_1$. 
This implies that $A_1^{\vec{x}, \vec{y}} \models \varphi_1(\vec{x}, \vec{y})$.

We want to prove that $A_1^{\vec{x}, \vec{y}} \models \varphi(\vec{x}, \vec{y})$. With this, we can argue that $\mathscr{A}_1 \Rightarrow \mathscr{A}$.
We prove by contradiction. 
Let $A_1^{\vec{x}, \vec{y}} \not\models \varphi(\vec{x}, \vec{y})$. Since $\varphi$ is a conjunction of various atomic formulas, there is at least one atomic formula that is not satisfied by $A_1^{\vec{x}, \vec{y}} $. 
\begin{itemize}
\item[-] Let $A_1^{\vec{x}, \vec{y}} \not\models$ $p \xrightarrow{e} q$ for $p, q \in \vec{x} \cup \vec{y}$. Let the node in $\mathscr{A}$ mapped to $p$ and $q$, through $M_f$ or $M_b$, be $(n_1, n_2)$ and $(n_1', n_2')$ respectively. From the definition of $M_b$ and $M_f$ we deduce that $n_1 \in N_1$ is mapped to $p$ and $n_1' \in N_1$ is mapped to $q$, through maps $M_{f1}$ and $M_{b1}$. 
Further, given the topological constraints on the alignment maps $NM$ and $EM$ one can show that $n_1 \xrightarrow{e} n_1'$ holds in $\mathscr{A}_1$.
Since $\varphi_1$ is obtained by conjoining all atomic node and edge predicates that holds in $\mathscr{A}_1$, it implies that $\varphi_1(\vec{x}, \vec{y}) \Rightarrow p \xrightarrow{e} q$. Since, $A_1^{\vec{x},\vec{y}}$ satisfies $\varphi_1(\vec{x}, \vec{y})$, it implies that $A_1^{\vec{x},\vec{y}} \models p \xrightarrow{e} q$, which is a contradiction.

\item[-] Let $A_1^{\vec{x}, \vec{y}} \not\models$ $\eta(p)$ for $p \in\vec{x} \cup \vec{y}$. Let the node in $\mathscr{A}$ mapped to $p$, through $M_f$ or $M_b$, be $(n_1, n_2)$. From the definition of $M_b$ and $M_f$ we deduce that $n_1 \in N_1$ is mapped to $p$ through maps $M_{f1}$ and $M_{b1}$. From the construction of $\mathscr{A}$, we know that $Lat(n_1) \sqsubseteq Lat(n_1, n_2)$. 
Since, $\varphi_1(\vec{x}, \vec{y})$ includes all atomic node predicates that hold over $n_1$, $\varphi_1(\vec{x}, \vec{y}) \Rightarrow [[Lat(n_1)]]$. Since $A_1^{\vec{x}, \vec{y}}$ satisfies $\varphi_1(\vec{x}, \vec{y})$, it implies that $A_1^{\vec{x}, \vec{y}}$ satisfies $Lat(n_1,n_2)$ and hence the formula $\eta(p)$, which is a contradiction.
\end{itemize}

Similarly, we show that $\mathscr{A}_2 \Rightarrow \mathscr{A}$. 
Together, this proves that $\mathscr{A}_1 \vee \mathscr{A}_2 \Rightarrow \mathscr{A}$.

\end{proof}

\noindent\textbf{Theorem} For any node mapping $NM \subseteq N_1 \times N_2$ and edge mapping $EM \subseteq E_1 \times E_2$ that is obtained from aligning $\mathscr{A}_1$ and $\mathscr{A}_2$, $\mathscr{A}_1 \vee \mathscr{A}_2 \Rightarrow \texttt{project}_{\{\mathscr{A}_1, \mathscr{A}_2\}}(\texttt{merge}_{(NM^\epsilon, EM^\epsilon)}(\mathscr{A}_1, \mathscr{A}_2))$.
\begin{proof}

Let $\mathscr{A}_1 = (N_1, E_1, Lat_1, M_{f1}, M_{b1})$ and $\mathscr{A}_2 = (N_2, E_2, Lat_2$, $M_{f2}, M_{b2})$ be the two UAPDGs, with the same set of free variables $\vec{x} = \{x_1, \cdots, x_n\}$.
Let the node and edge mappings obtained on aligning $\mathscr{A}_1$ and $\mathscr{A}_2$ be $NM$ and $EM$ respectively. 
The ILP constraints ensure that $NM$ respect the maps $M_{f1}$ and $M_{f2}$, i.e., $M_{f1}(n_1) = M_{f2}(n_2) \Rightarrow (n_1, n_2) \in NM$.

Let $N_1' =\{n_1 \mid (n_1, n_2) \in NM\}$ and  $E_1'  =\{e_1 \mid (e_1, e_2) \in EM\}$ be subsets of $N_1$ and $E_1$ respectively obtained by projecting maps $NM$ and $EM$ to $\mathscr{A}_1$. We define sets $N_2'$ and $E_2'$ by projecting $NM$ and $EM$ to $\mathscr{A}_2$ in the same way. 
Let $\mathscr{A}_1' = \texttt{project}(\mathscr{A}_1) \downarrow_{(N_1', E_1')}$ and  $\mathscr{A}_2' = \texttt{project}(\mathscr{A}_2) \downarrow_{(N_2', E_2')}$. 
From Lemma~\ref{lemma:projection}, we know that $\mathscr{A}_1 \Rightarrow \mathscr{A}_1'$ and $\mathscr{A}_2 \Rightarrow \mathscr{A}_2'$. 
If we can prove that $A_1' \vee A_2' \Rightarrow \texttt{merge}_{(NM, EM)}(\mathscr{A})$ then we can easily show that $A_1 \vee A_2 \Rightarrow \texttt{merge}_{(NM, EM)}(\mathscr{A})$. 

Let us therefore prove that $A_1' \vee A_2' \Rightarrow $ $\texttt{project}($ $\texttt{merge}_{(NM, EM)}($ $\mathscr{A}))$. Without loss of generality, let us assume that bound variables in $\mathscr{A}_1'$ and $\mathscr{A}_2'$ are labeled such that 
$M_{b1}(n_1) = M_{b2}(n_2) \Rightarrow (n_1,n_2) \in NM$. Since, 
\begin{itemize}
\item[(a)] all nodes in $N_1'$ and $N_2'$ and all edges in $E_1'$ and $E_2'$ are mapped through $NM$ and $EM$ respectively (from construction of $\mathcal{A}_1'$ and $\mathcal{A}_12'$)
\item[(b)] If $(n_1, n_2) \in NM$, then if $n_1$ is mapped to a free variable,it follows from the properties of $NM$ that $M_{f1}(n_1) =M_{f2}(n_2)$. On the other hand, if $n_1$ is mapped to a bound variable, then it follows from our assumption above that $M_{b1}(n_1) = M_{b2}(n_2)$. 
\end{itemize}
it follows from Lemma~\ref{lemma:restricted_soundness} that $A_1' \vee A_2' \Rightarrow \texttt{merge}_{(NM, EM)}(\mathscr{A})$. 
From the definition of the $\texttt{project}$ operator, one can easily see that UAPDG $\texttt{merge}_{(NM, EM)}(\mathscr{A}) = \texttt{project}_{\{\mathscr{A}_1, \mathscr{A}_2\}}($ $\texttt{merge}_{(NM^\epsilon, EM^\epsilon)}($ $\mathscr{A}))$. This concludes the proof of the theorem.

\end{proof}

\section{Proof of Theorem~\ref{thm:overall_soundness}}
\label{app:proof2}

\textbf{Theorem}~[Soundness of Rule Synthesis:]
Given a set of violating examples $\mathscr{V} = \{V_1, \cdots, V_m\}$ and conforming  examples $\mathscr{C} = \{C_1, \cdots, C_n\}$, if the algorithm \verb+synthesizeRule+ successfully returns a return $R$, then $V_i \models R$ and $C_i \not\models R$.

\begin{proof}
It is easy to see in the algorithm \verb+synthesizeRule+ (refer to Algorithm~\ref{alg:overall}) that a rule $R$ is successfully returned only if $V_i \models R$, for all $i \in \{1,\cdots, m\}$.
We now argue that when $R$ is successfully returned, $C_i \not\models R$, for all $i \in \{1,\cdots, n\}$. Proving this involves two cases:

\noindent(1) When $C_i \not\models \exists \vec{x}.\textit{pre}(\vec{x})$. Since $R$ strengthens the precondition, it follows that $C_i \not\models R$.

\noindent  (2) When $C_i^{\vec{x}} \models \textit{pre}(\vec{x})$. Since $R = \exists \vec{x}. \textit{pre}(\vec{x}) \wedge \neg \exists \vec{y}. \textit{post}(\vec{x}, \vec{y})$, to prove that $C_i^{\vec{x}} \not\models R$, we need to prove that $C_i^{\vec{x}} \not\models \neg \exists \vec{y}. \textit{post}(\vec{x}, \vec{y})$, or in other words, $C_i^{\vec{x}} \models \exists \vec{y}. \textit{post}(\vec{x}, \vec{y})$. 

\medskip\noindent Proof that $C_i^{\vec{x}} \models \exists \vec{y}. \textit{post}(\vec{x}, \vec{y})$: 
Let $\mathscr{C}_1^{\vec{x}}, \cdots, \mathscr{C}_k^{\vec{x}}$ be a partition in the partition list in method \texttt{\small synthesizePC}. One can argue that $\mathscr{C}_1^{\vec{x}} \cup \cdots \cup \mathscr{C}_k^{\vec{x}} = \mathscr{C'}^{\vec{x}}$ where $\mathscr{C'}^{\vec{x}}$ is the set of valuated conforming examples that satisfy the precondition and is computed at line 2 in algorithm~\ref{alg:overall}. Also note that since $C_i^{\vec{x}} \models \textit{pre}(x)$, it follows from the definition of $\mathscr{C'}^{\vec{x}}$ that  $C_i^{\vec{x}} \in  \mathscr{C'}^{\vec{x}}$. Without loss of generality, let us assume that $C_i^{\vec{x}} \in \mathscr{C}_j^{\vec{x}}$ (the item in $j^{th}$ partition). 
As \texttt{\small getConjunctiveRule} returns the UAPDG obtained on \texttt{\small merging} all its input Valuated PDGs, it follows from the soudness of \texttt{\small merge} algorithm (Theorem~\ref{thm:merge_soundness}) that $C_i^{\vec{x}} \models \exists\vec{y}. \textit{post}_j(\vec{x}, \vec{y})$.	 Consequently, $C_i^{\vec{x}} \models \exists\vec{y}. \textit{post}(\vec{x}, \vec{y})$.

\end{proof}

\section{Synthesized Rule Examples}
\label{app:apdg}

In this appendix, we list an example code change and visualize the synthesized precondition and postcondition UAPDGs for few code-quality rules described in Table~\ref{fig:bp_description}. Refer to the supplementary material for details about all synthesized rules.
 
\clearpage

\begin{figure*}[h!t!]
\centering
\caption{Rule \texttt{\small check-actionbar}:
The method `getSupportActionBar` returns `null` if the Android activity does not have an action bar. One must null-check the
value returned by `getSupportActionBar`, if the action bar is not explicitly set by a `setSupportActionBar` call.\\
(a) Example code change
(b) Precondition of the synthesized rule. (c) A disjunct in the synthesized postcondition: if `setSupportActionBar` is explicitly called, then the example is conforming. Note this code variation for conforming examples is not present in the code change and included in the rule during rule refinement. (d) A disjunct in the synthesized postcondition: the value returned by `getSupportActionBar` must be null checked.}
\vspace{1cm}

(a)
\begin{minipage}{0.7\textwidth}
  \includegraphics[width=\textwidth]{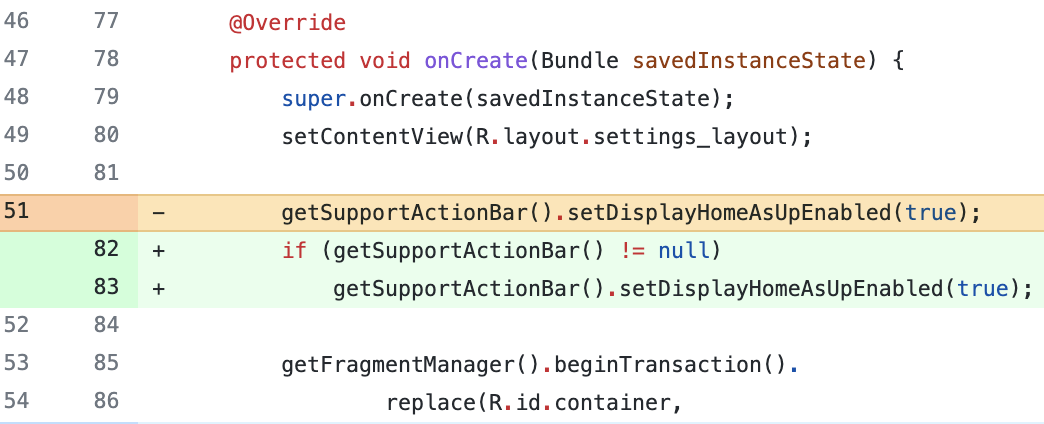}
\end{minipage}
\vspace{1cm}

(b)
\begin{minipage}{0.98\textwidth}
  \includegraphics[width=\textwidth]{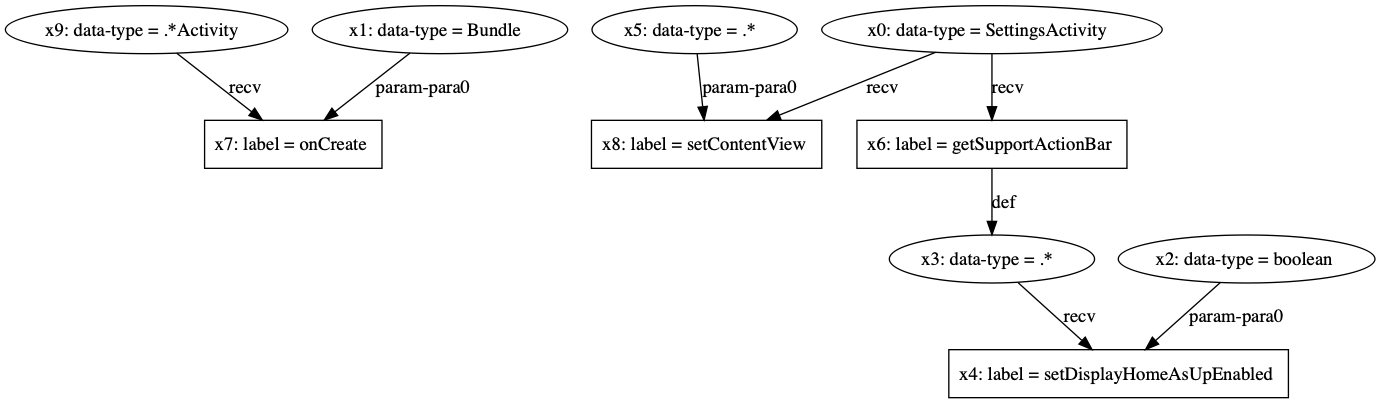}
\end{minipage}
\vspace{1cm}

(c)
\begin{minipage}{0.98\textwidth}
  \includegraphics[width=\textwidth]{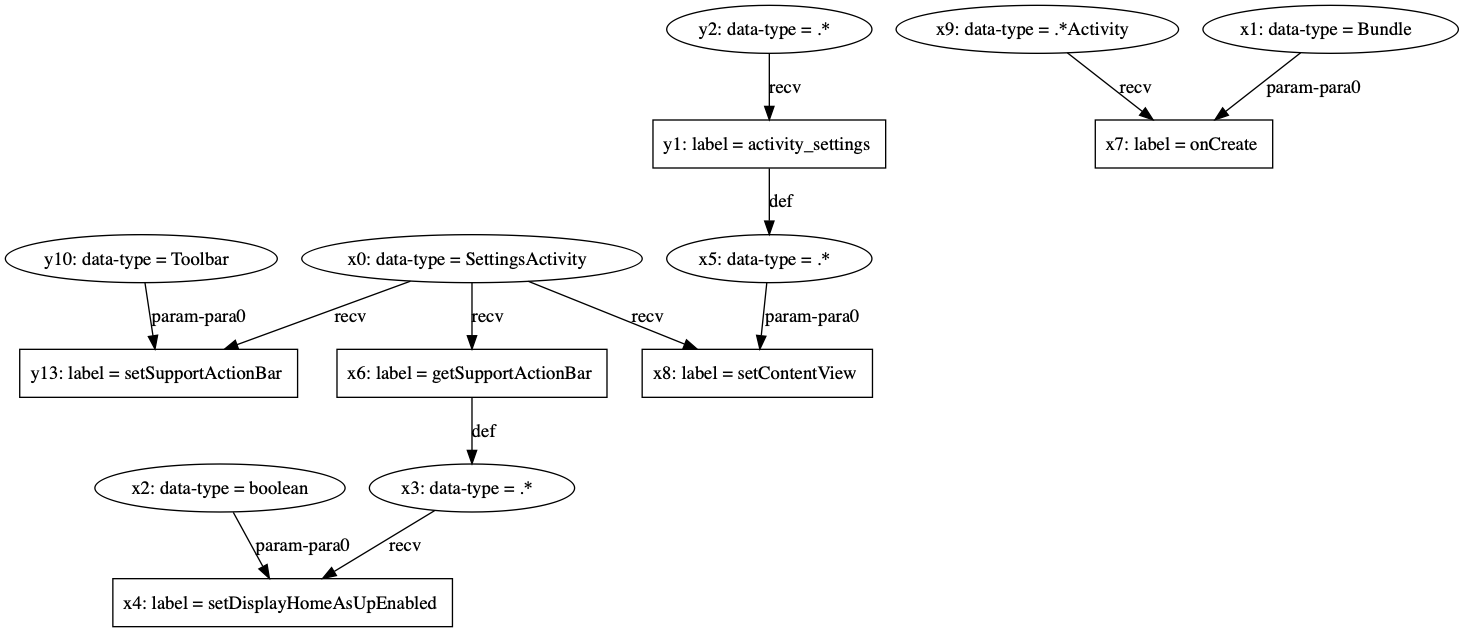}
\end{minipage}
\end{figure*}

\begin{figure*}[h!t!]
\centering
(d)
\begin{minipage}{0.98\textwidth}
  \includegraphics[width=\textwidth]{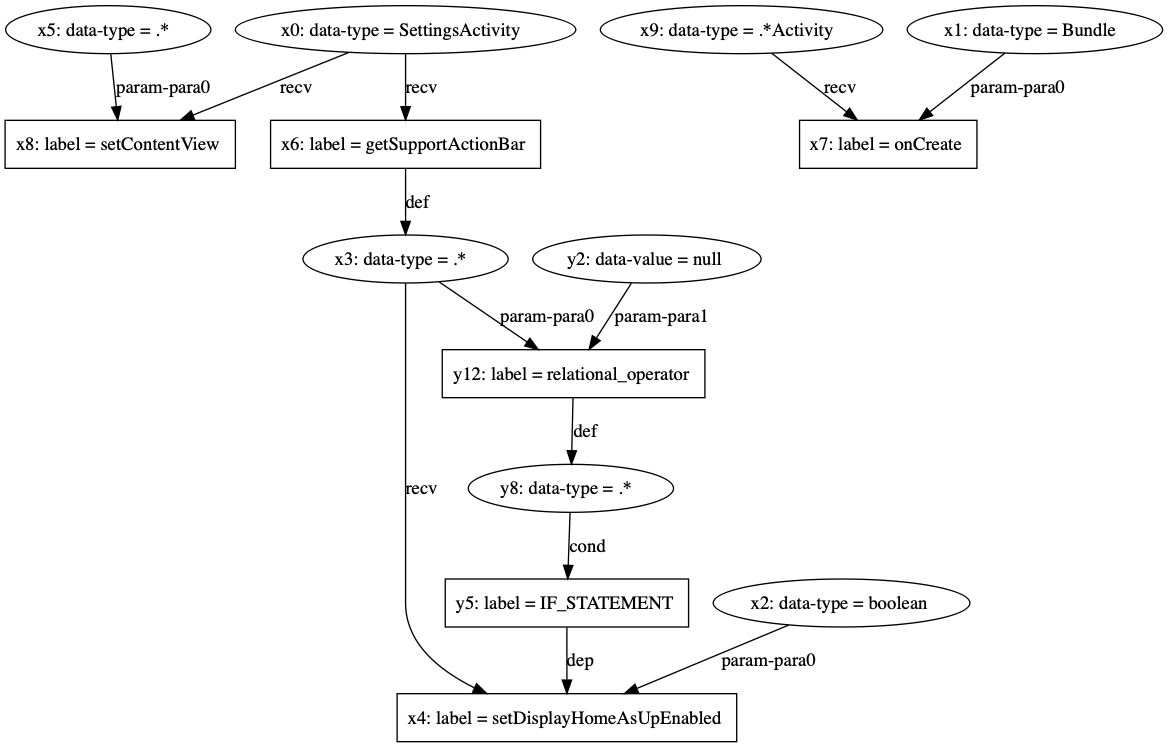}
\end{minipage}
\end{figure*}

\clearpage

\begin{figure*}[h!t!]
\centering
\caption{Rule \texttt{\small upgrade-http-client}:
The rule detects code that calls `HttpClient.executeMethod`. This method has been replaced in the HttpClient library version 4 with 
method `execute`. One must upgrade to the latest version of the HttpClient library.\\
\textit{URL: \href{https://stackoverflow.com/questions/40795037}{https://stackoverflow.com/questions/40795037}}\\
(a) Example code change 
(b) Precondition of the synthesized rule }
\vspace{1cm}

(a)
\begin{minipage}{.9\textwidth}
  \includegraphics[width=\textwidth]{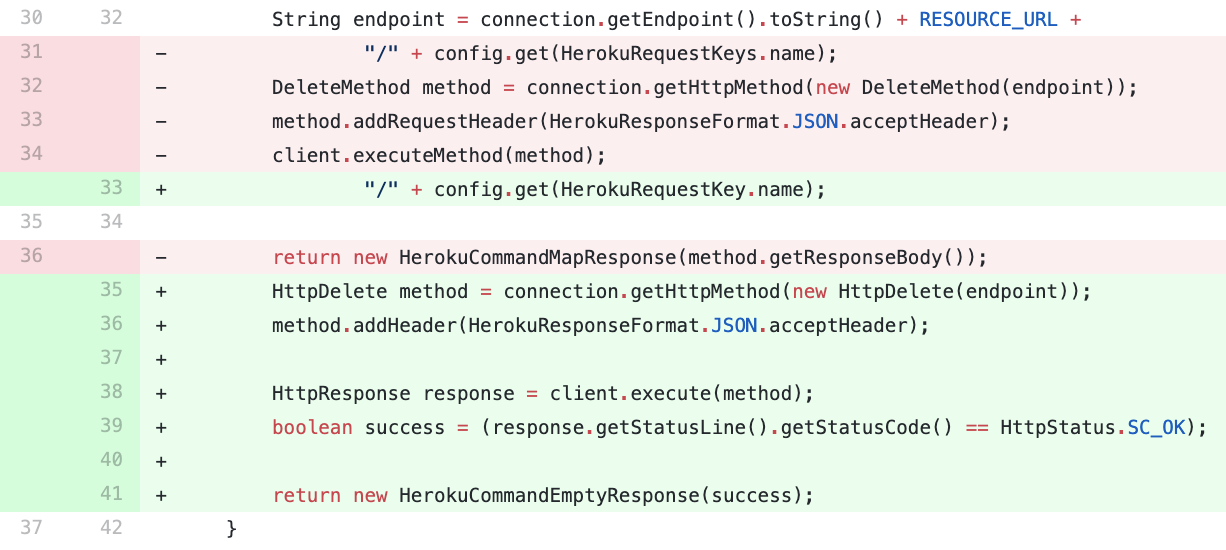}
\end{minipage}
\vspace{1cm}

(b)
\begin{minipage}{.6\textwidth}
  \includegraphics[width=\textwidth]{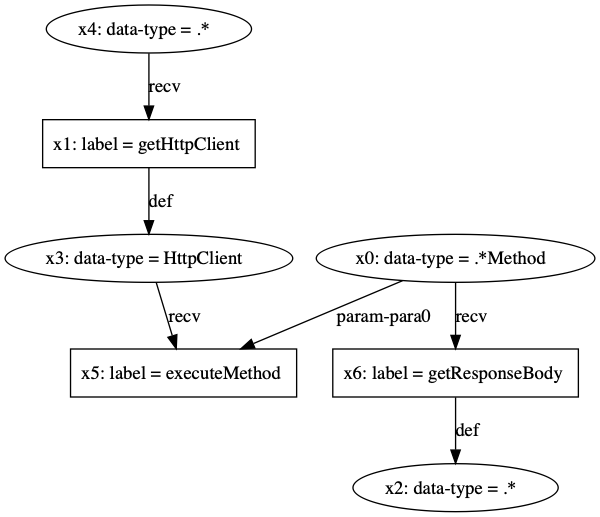}  
\end{minipage}
\end{figure*}

\clearpage

\begin{figure*}[h!t!]
\centering
\caption{
Rule \texttt{\small start-activity}:
When launching an Android activity with `startActivity(Intent, ...)`, one must  check `Intent.resolveActivity(..)` for null. This checks 
if there exists an app on the   device that can receive the implicit intent and launch the activity.  Otherwise, the  app will crash  when 
`startActivity` is called.  This is not required when the activity  is part of the same app,  or when `startActivity` is called within a try-catch block.\\
\textit{URL: \href{https://developer.android.com/guide/components/intents-common}{https://developer.android.com/guide/components/intents-common}} \\
(a) Example code change 
(b) Precondition of the synthesized rule 
(c) A disjunct in the synthesized postcondition: if `Intent` is created from a class in the same application, then the example is conforming. Note this code variation for conforming examples is not present in the code change and included in the rule during rule refinement. (d) A disjunct in the synthesized postcondition: if a catch block exists to handle an exception thrown by `startActivity`, the code is conforming.  This code vairation is also learned from conforming examples during rule refinement.  (e) A disjunct in the synthesized postcondition: if `Intent.resolveActivity(..)` is checked for null, the code is conforming. 
}
\vspace{0.3cm}

(a)
\begin{minipage}{.95\textwidth}
   \includegraphics[width=\textwidth]{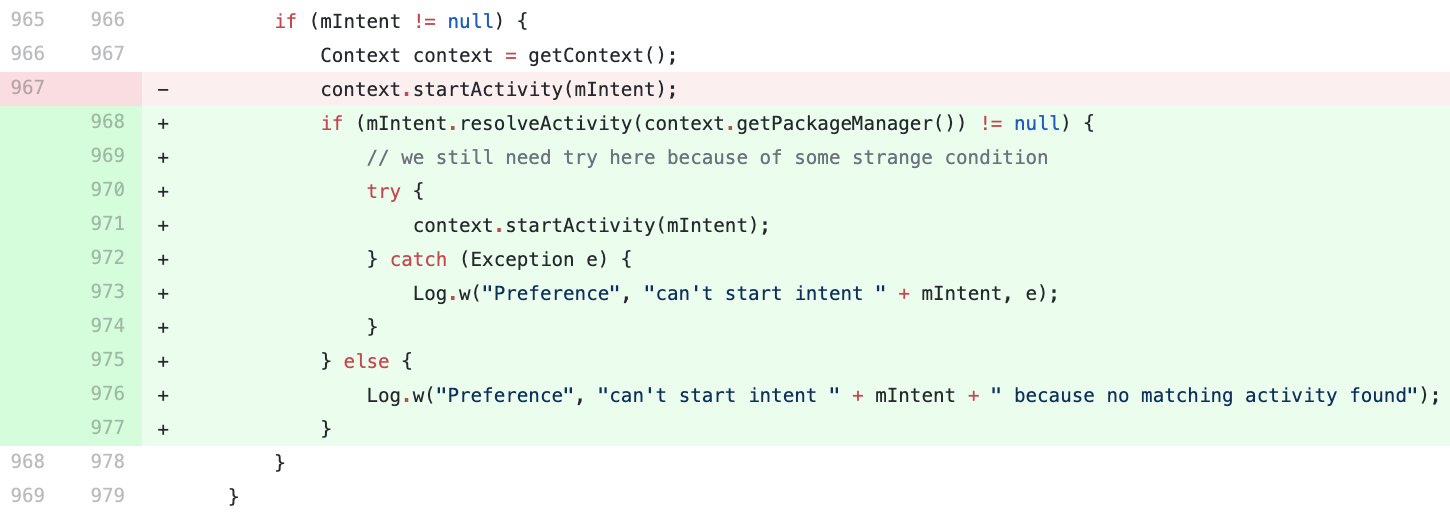}
\end{minipage}
\vspace{0.3cm}

(b)
\begin{minipage}{.45\textwidth}
   \includegraphics[width=\textwidth]{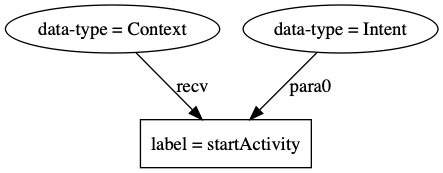}
\end{minipage}
\vspace{1cm}

(c)
\begin{minipage}{0.5\textwidth}
   \includegraphics[width=\textwidth]{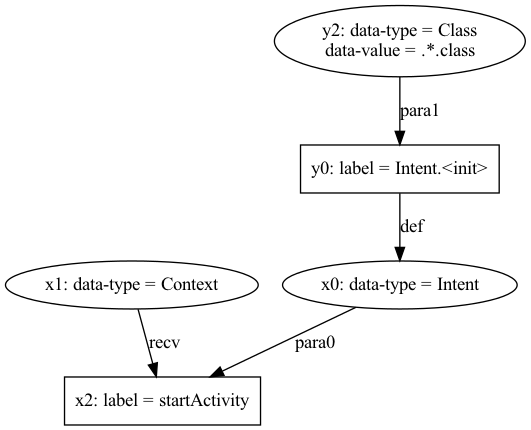}
\end{minipage}
\end{figure*}

\begin{figure*}[h!t!]
\centering
(d)
\begin{minipage}{0.85\textwidth}
   \includegraphics[width=\textwidth]{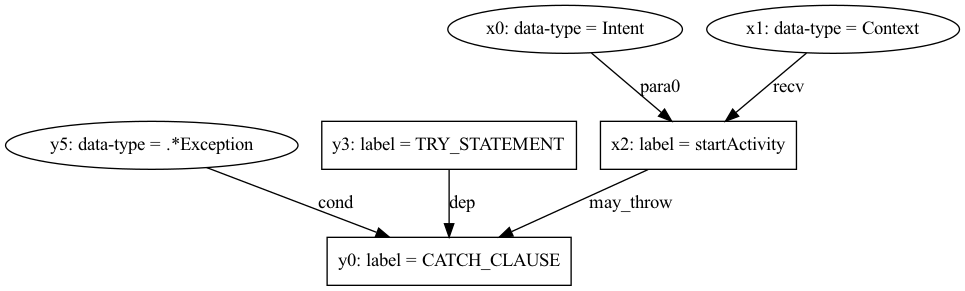}
\end{minipage}
\vspace{1cm}

(e)
\begin{minipage}{0.55\textwidth}
   \includegraphics[width=\textwidth]{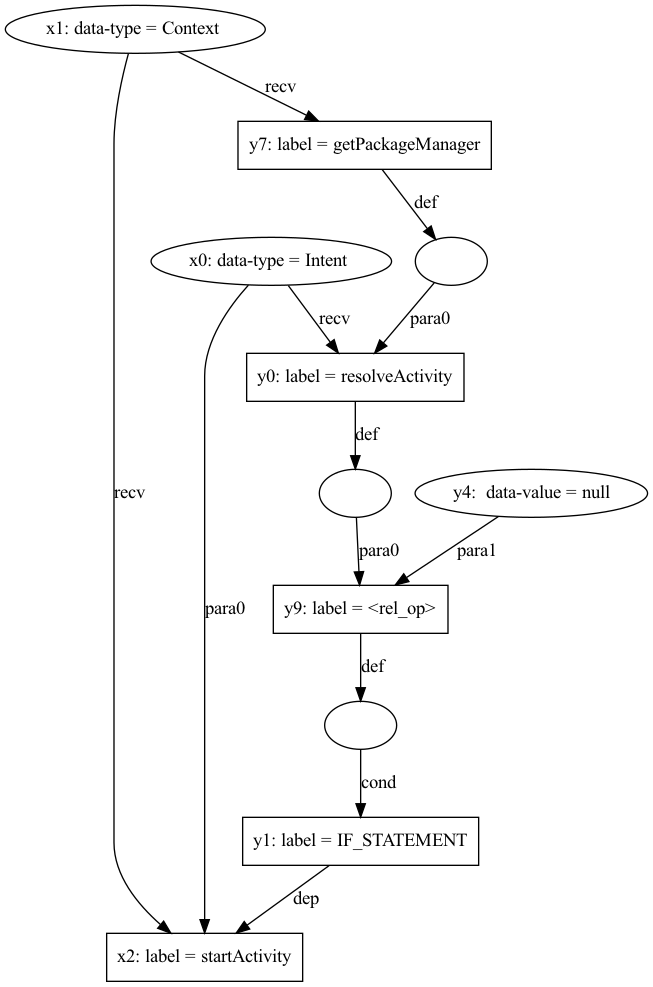}
\end{minipage}

\end{figure*}

\clearpage

\begin{figure*}[h!t!]
\centering
\caption{
Rule \texttt{\small deserialize-json-array}:
The rule detects code that deserializes a list of JSON items by iterating in a loop. Instead, one can directly deserialize into a list by 
specifying the correct parameterized type using the `TypeToken` class.\\
\textit{URL: \href{https://github.com/google/gson/blob/master/UserGuide.md\#TOC-Serializing-and-Deserializing-Generic-Types}{https://github.com/google/gson/blob/master/UserGuide.md\#TOC-Serializing-and-Deserializing-Generic-Types}}\\
(a) Example code change 
(b) Precondition of the synthesized rule}
\vspace{0.3cm}

(a)
\begin{minipage}{1.1\textwidth}
   \includegraphics[width=\textwidth]{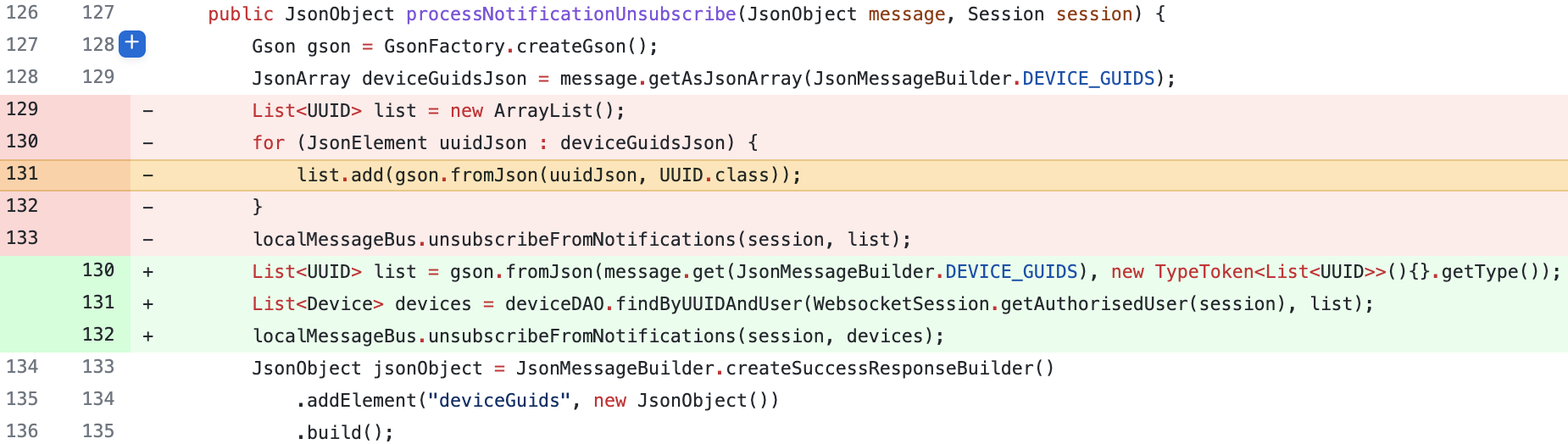}
\end{minipage}
\vspace{1cm}

(b)
\begin{minipage}{0.95\textwidth}
   \includegraphics[width=\textwidth]{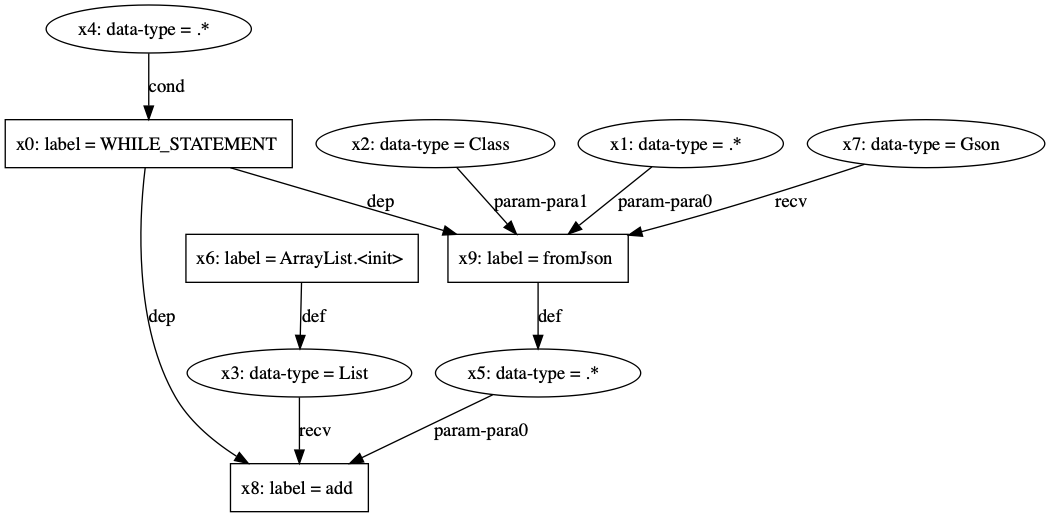}
\end{minipage}
\vspace{0.3cm}

\end{figure*}

\end{document}